\documentclass[preprint,12pt]{elsarticle}

\usepackage{graphicx} \usepackage{amsmath} \usepackage{amsfonts}
\usepackage{amssymb}
\usepackage{comment}
\usepackage{wrapfig}
\usepackage{floatrow}
\usepackage{breqn}
\usepackage{bm}

\usepackage{epstopdf}
\DeclareFloatFont{cusize}{\normalsize}
\def\deg{\circ}
\def\fpt{\emph{findpts}}

\def\dO{\partial \Omega}

\def\tOmega{\tilde {\Omega}}

\def\tOmega{{\tilde{\Omega}}}

\def\scriptO{{{\it O}\kern -.42em {\it `}\kern + .20em}}
\def\RR{{{\rm l}\kern - .15em {\rm R} }}
\def\PP{{{\rm l}\kern - .15em {\rm P} }}
\def\L2{{{\sf L}^2}}
\def\H1{{{\sf H}^1}}
\def\PN2{{\PP_{N}-\PP_{N-2}}}

\def\complex{{{\rm C} \kern - .53em {\rm l} \kern + .38em}}

\def\a1{{ | \lambda_{\min} |}}

\def\l1{{   \lambda_{\min}  }}

\def\bu0{{\underline {\bf 0}}}

\def\bu{{\bf u}}

\def\bx{{\bf x}}

\def\bq{{\bf q}}

\def\bxi{{\bm{\xi}}}

\def\u0{{\underline 0}}

\journal{Computers \& Fluids}

\begin{document}

\begin{frontmatter}

\title{Direct Numerical Simulation of Rotating Ellipsoidal Particles using Moving Nonconforming Schwarz-Spectral Element Method}

\author{Ketan Mittal\fnref{label1}}
\author{Som Dutta\fnref{label2}}
\author{Paul Fischer\fnref{label1,label3}}
\fntext[label1]{Mechanical Science \& Engineering, University of Illinois at Urbana-Champaign, 1206 W. Green St., Urbana, IL 61801}
\fntext[label2]{Mechanical \& Aerospace Engineering, Utah State University, 4130 Old Main Hill, Logan, UT 84332}
\fntext[label3]{Computer Science, University of Illinois at Urbana-Champaign, 201 N. Goodwin Ave., Urbana, IL 61801}

\begin{abstract}

We present application of a highly-scalable overlapping grid-based nonconforming Schwarz-spectral element method (Schwarz-SEM) to study the dynamics of rotating ellipsoidal particles.  The current study is one of the first to explore the effect of rotation on ellipsoidal particles using direct numerical simulation.  The rotating ellipsoidal particles show substantial difference in the dynamics of the flow, when compared against non-rotating particles. This difference is primarily due to periodic attachment and separation of the flow to the surface of the particle for the rotating cases, which results in a higher drag on the particle when compared to the corresponding non-rotating cases. The dynamics is also different from a rotating spherical particle, where a steady shear layer develops near the surface of the sphere.  For the rotating ellipsoidal particles, we observe that there is a phase-difference between the position of observed maximum and minimum drag, and the position of expected maximum and minimum drag (i.e., maximum and minimum projected area). A similar phase-difference is also observed for the lift acting on the rotating ellipsoidal particles.  The results presented here demonstrate the importance of explicitly modeling the shape and rotation of particles for studying dynamics of non-spherical particles. Finally, this study also validates the use of non-conforming Schwarz-SEM for tackling problems in fully resolved particulate flow dynamics.
\end{abstract}

\end{frontmatter}

\section{Introduction}\label{intro}

Particle-laden flows are common in natural and engineered systems such as sediment transport in rivers and oceans \cite{dutta2017three,dutta2016,dutta2014}, tephra dispersal through volcanic plumes \cite{ward2019}, transport in industrial systems during chemical processing and in oil pipelines \cite{li2001multiscale}, and the human respiratory system \cite{kleinstreuer2010airflow}.
Particles are often assumed to be spherical when their transport in the above stated systems is modeled.
Though, this assumption has been found to be invalid for particles going through different natural and industrial processes \cite{voth2017anisotropic}.
Nonspherical particle shapes in nature and industry are driven by design needs \cite{lundell2011fluid,erni2009}, different generation processes \cite{best1998influence,moffet2009}, and evolutionary natural selection \cite{sabban2011,pedley1992}.
Recent studies have explored the
effect of particle-shape on dynamics of the particle, especially the effect of non-sphericity of the particle on its movement \cite{voth2017anisotropic,njobuenwu2014,van2015}.
In the process of studying non-spherical particles, irregular shaped particles are often idealized to ellipsoids, cylinders or cuboids \cite{ouchene2015drag}.
In the current paper, our goal is to understand how the combined effect of the shape of a particle and its rotation impact the flow around the particle, and the forces experienced by it.

Historically, studies on dynamics of nonspherical particles have usually focused on the drag force acting on the particle. In that context, Chhabra et al. \cite{chhabra1999drag} have compiled a definitive list of drag coefficient relationships, and defined two approaches for drag coefficient relationships.  The first approach does not account for the shape or orientation of the particle \cite{tran2004,yow2005}, and the second approach accounts for only the orientation of the particle. The second approach has led to correlations that were established using direct numerical simulation (DNS) based on the immersed boundary method \cite{rosendahl2000}. In contrast, Holzer and Sommerfeld \cite{holzer2008} have taken an approach in which they account for the shape of the ellipsoid using sphericity and crosswise sphericity, but do not account for orientation of the particle.

The importance of orientation with respect to the background flow for ellipsoidal (or any nonspherical) particles is underlined by the fact that particle-transport in most natural and industrial systems is a combination of particle translation and rotation \cite{voth2017anisotropic,dobson2014flow}.  Particles in these systems could rotate due to collision with other particles and the walls, or due to presence of high vorticity and mean shear of the fluid. Thus, it is crucial to model the shape and rotation of the particle for studying particle-laden flows.

Recently Zastawny et al. \cite{zastawny2012} have conducted DNS of nonrotating nonspherical particles at different orientations, and DNS of rotating ellipsoids using an immerse boundary method. In \cite{zastawny2012}, the authors show that the orientation of nonspherical particles can have a significant impact on the drag and lift coefficient of the particle, and also describe how the torque coefficient for rotating nonspherical particles varies with the rotational Reynolds number. Zastawny et al. compared their results against theoretical expression derived under the Stokes flow assumption ($Re << 1$) by Happel and Brenner \cite{happel1965}.  Incidentally, the expression by Happel and Brenner has been used extensively to study motion of inertial ellipsoidal particles \cite{mortensen2008,marchioli2010}, though recent studies have questioned its suitability, especially for cases that has particle Reynolds number ($Re$) greater than 10 \cite{ouchene2015,ouchene2016}. Additionally, Zastawny et al. do not sufficiently describe how the shape and rotation of nonspherical particle impacts the flow around the particle.

Ouchene et al. \cite{ouchene2015,ouchene2016} have also conducted DNS of flow over ellipsoidal particles of different aspect ratios, at varying incident angles (orientations). In \cite{ouchene2016}, the authors have proposed correlations for drag and lift coefficients for a large range of Reynolds number and aspect ratios.  Despite being a step in the right direction, their approach does not account for the dynamics induced by rotation of the particle, e.g.  the \textit{Magnus-Robins} effect on a rotating spherical body/particle that results in an additional force on it \cite{batchelor}.  Experiments have also shown the importance of particle rotation in enhancing turbulence in flow at even moderate Reynolds number \cite{best1998influence}, further underlining the need to capture the effect of particle rotation on the flow.

One of the first systematic study for capturing the impact of particle rotation was done by Rubinow and Keller \cite{rubinow1961}, who used the \textit{Stokes-Oseen} expansion to derive an approximate expression for lift on a rotating sphere  ($F_L = \pi D^3 \rho \tilde{\Omega} U_{\infty}/8$) that is valid for flow in the Stokes regime ($Re = U_{\infty}D/\nu << 1$). Here, $D$ is diameter of the sphere, $U_{\infty}$ is the free-stream (background) velocity, $\tilde{\Omega}$ is the angular velocity of the sphere, $\rho$ is density of the fluid, and $\nu$ is the kinematic viscosity of the fluid. Tsuji et al.  \cite{tsuji1985} approximated the relationship $C_L = 0.4 \pm 0.1 \Omega^{*}$ between lift coefficient ($C_L$) and nondimensional rate of rotation ($\Omega^{*} = \tilde{\Omega} D/2 U_{\infty}$) using experimental data valid for $\Omega^{*} \leq 0.7$ and $550 \leq Re \leq 1600$.
 Loth \cite{loth2008} approximated a relationship for $C_L$ for a rotating sphere that correlated it with $Re$ and $\Omega^{*}$ as:
 \begin{eqnarray}
 \nonumber
 C_L = \Omega^{*} \bigg( 1 - \left\{0.675 + 0.15 \left( 1 + \tanh \left[ 0.28( \Omega^{*} - 2) \right] \right\} \right) \bigg) \bigg(\tanh\left[0.18Re^{0.5}\right]\bigg).
 \end{eqnarray}
 With the increase of computing power, fully resolved simulations (DNS and high-resolution LES) have become the primary tool to study  the dynamics of rotating particles.  Dobson et al., Kim et al., and Poon et al. have done high-resolution LES and DNS calculations to study flow around a rotating sphere for a range of Reynolds number and rotation rates \cite{dobson2014flow,kim2009laminar,poon2010laminar}.

A survey of the literature shows that there exists abundant data for flow past
rotating spherical particles and for flow past static nonspherical particles at
different orientations. However, there is a need for a systematic study that
captures the impact of the shape and rotation of nonspherical particles on the
flow around them. In this paper, we conduct DNS using a high-order
overlapping grid-based spectral element method (SEM) to demonstrate several new results.
 First, we demonstrate that the shape and rotation of the particle significantly impacts the
drag and lift forces experienced by the particle (e.g., the maximum drag force
experienced by a rotating ellipsoid is more than twice the drag force
experienced by a rotating sphere). Second, the rotation of the particle leads
to a phase difference between the orientation at which a particle experiences
maximum drag and the orientation at which the frontal area is maximum.
This phase difference between the two orientation is due to the interaction of the background flow with the attached
high-speed region moving from the leeward side to the windward side of the
particle. Finally, we also demonstrate that explicitly modeling the particle
rotation is essential for accurately capturing the impact of the particle on
the flow around it. We note that unlike some of the work in existing
literature, we do not seek to sweep a wide range of particle sizes and
rotational rates (or orientations) to develop correlations for drag and lift
coefficients. Instead, our goal is to systematically understand how the shape
and rotation of the particle impacts the flow around it, to accurately model
thousands of arbitrary shaped particles in particle-laden flows.

Henceforth, the paper is divided into four major sections. In Section 2, the
numerical method and the setup for the simulations has been described,
particularly focusing on the nonconforming Schwarz-spectral element
method (Schwarz-SEM) framework. In Section 3 the simulation results are
discussed, where we first validate our framework using existing DNS for a rotating
sphere \cite{dobson2014flow}, and then illustrate the results from the
rotating ellipsoidal cases. In Section 4, the combined effect of rotation and
shape of the ellipsoid on the drag and lift is discussed and compared with
several nonrotating configurations. Finally, the paper ends with a summary of
the findings and potential avenues for future research.

\section{Methodology}
\subsection{Governing equations \& problem setup}
To understand the impact of the particle shape and rotation on the interaction
of the particle with the background flow, we solve the incompressible
Navier-Stokes equations (INSE) using the spectral element method. The nondimensional
constant-density incompressible Navier-Stokes equations in a given
computational domain $\Omega(t)$ in $\RR^d$ at time $t$ are given by
\begin{eqnarray}
\label{eq:nsemomentum}
 \quad \frac{\partial \bu}{\partial t}  + \bu\cdot\nabla{\bu} = - \nabla{p} +
\frac{1}{Re} \nabla^2{\bu}, \\
\label{eq:nsemass}
\nabla\cdot \bu = 0,
\end{eqnarray}
where $\bu(\bx,t)$ and $p(\bx,t)$ represent the velocity and pressure solution
as a function of position $\bx \in \RR^d$ and time $t$, and $Re=UL/\nu$ is the
Reynolds number based on a velocity scale ($U$), length scale ($L$), and the
kinematic viscosity ($\nu$). The solution of the INSE also depends on the
initial and boundary conditions, which we will discuss for our problem
later.  We note that the solution $\bu(\bx,t)$ and $p(\bx,t)$ (and
other functions such as the mesh velocity) are a function of position and time,
but we use $\bu$ and $p$ for brevity.

\begin{figure}[t!] \begin{center}
$\begin{array}{c}
\includegraphics[height=60mm]{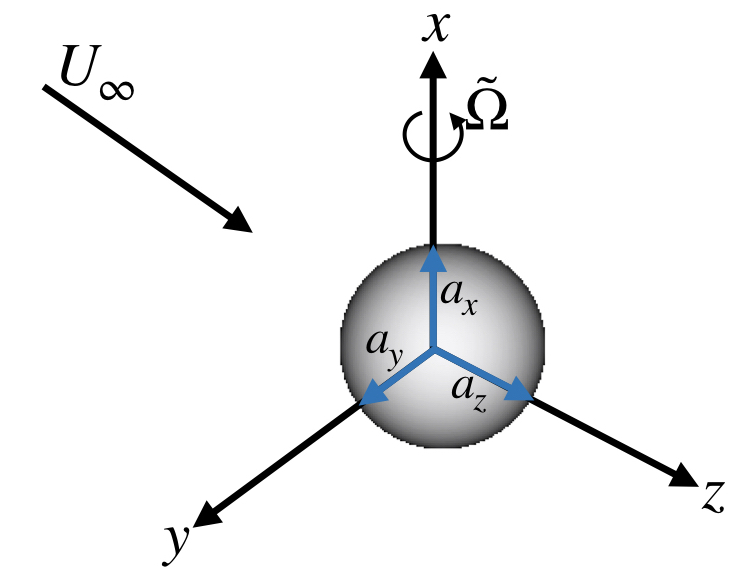}
\end{array}$
\end{center}
\caption{Schematic showing the direction of the
rotation of particle with respect to the flow.}
\label{fig:schematic}
\end{figure}

Following Dobson \cite{dobson2014flow}, we setup the numerical simulations such
that the rotation of axis of the particle is normal to the direction of the
flow. Figure \ref{fig:schematic} shows that the background flow ($U_{\infty}$)
is along the $z$-axis, and the particle rotates at a constant angular velocity
($\tOmega$) around the $x$-axis.  In Fig.  \ref{fig:schematic}, $a_j$
represents the length of the principal axis of the sphere in $j$th direction,
and $a_x=a_y=a_z$ for a sphere. The Reynolds number of the flow is $Re =
U_{\infty}D/\nu$, where the length scale is
the particle diameter ($D=2a_z$). The
nondimensional rotation rate of the particle is determined as
$\Omega^*=0.5\tilde{\Omega}D/U_{\infty}$.

We note that we set $a_x=a_y=a_z=0.5D$ for the sphere, and to
understand how the shape of particle impacts the dynamics of the flow around
it, we consider two different cases: $a_y=0.25D$ and $0.75D$.  Additionally, in
order to validate our methodology, we consider the case with $Re=300$ and
$\Omega^*=1$ for a rotating spherical particle, which has been studied by
Dobson et al. \cite{dobson2014flow}, Kim et al. \cite{kim2009laminar}, and Poon et al. \cite{poon2010laminar}.

\subsection{Domain decomposition} For a rotating spherical particle, the domain
can be modeled with a single
static conforming mesh with Dirichlet conditions imposed on the surface of the
sphere to model the effect of rotation. For arbitrary shaped nonspherical
particles however, this approach is not straightforward.  Using a single static
mesh is also not feasible for our target application where we will model
hundreds of arbitrary shaped particles, each of whose rotation depends on the
dynamics of the flow around it. With this target problem in mind, we model the
domain with a single rotating particle using overlapping meshes with the
Schwarz-SEM framework. The advantage of this approach is that it
simplifies mesh generation and validates our method against a flow with complex
flow structures.

\begin{figure}[t!] \begin{center}
$\begin{array}{c}
\includegraphics[height=60mm]{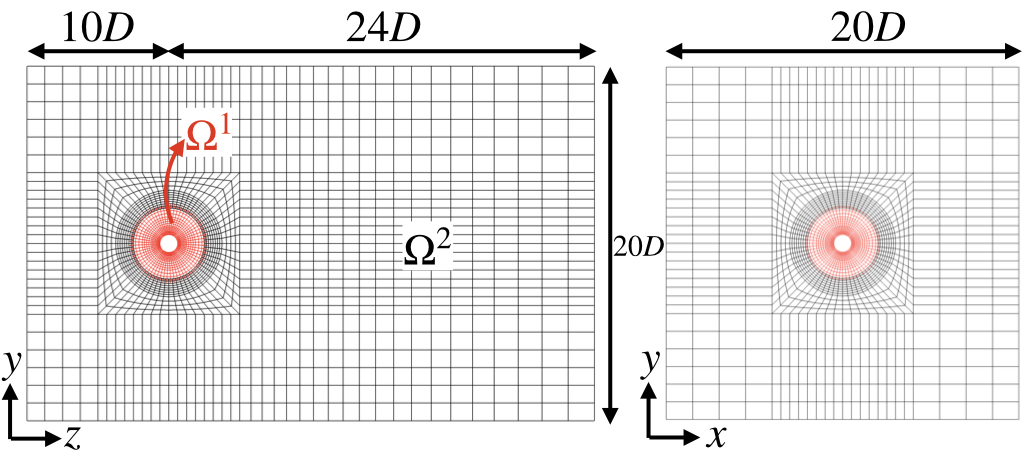}
\end{array}$
\end{center}
\caption{Slice view of the rotating mesh used for modeling the rotating particle (red),
and the static background mesh (black).}
\label{fig:basemesh}
\end{figure}

The Schwarz-SEM framework is based on the overlapping Schwarz (OS) method
\cite{schwarz1870,smith2004domain} for solving the incompressible Navier-Stokes
equations using the spectral element method.
For the flow
past a rotating particle, we partition the domain into $S=2$ overlapping
subsets; a rotating interior mesh ($\Omega^1$ with $E=24,576$ spectral
elements) captures the flow around the particle, which is overlapped with a
static background mesh ($\Omega^2$ with $E=41,216$), shown in Fig.
\ref{fig:basemesh}. Figure \ref{fig:basemesh} also indicates the domain
dimensions in terms of the diameter of the particle ($D=2a_z$). The radial
extent of the inner mesh is $2.25D$ and the overlap width between the two
subdomains is $0.25D$. The background mesh is periodic in $x$- and
$y$-direction, and uniform inflow and outflow boundary conditions are imposed
in the $z$-direction. Additionally, the inner mesh has
a moving boundary condition for the surface of the particle.

We note that in addition to simplifying mesh generation for arbitrary shaped particles,
the Schwarz-SEM framework allows use of
spatial resolution in each mesh based on the physics of the flow in that
region. Here, since we can anticipate that there will be relatively fine scale
structures in the wake behind the particle, the mesh is denser in that region
as compared to everywhere else in the domain.

\subsection{Schwarz-SEM framework}

In the Schwarz-SEM framework, we use the Arbitrary Lagrangian-Eulerian (ALE)
formulation \cite{donea2004arbitrary} for representing the solution of the
incompressible Navier-Stokes equations in overlapping moving domains (meshes)
with the spectral element method (SEM).

The spectral element method (SEM) is a high-order weighted residual method that
was introduced by Patera \cite{patera84} and has been used to solve a variety
of challenging fluid dynamics and heat transfer problems
\cite{dutta2016,merzari2017,vinuesa2018turbulent}. The basis functions in
the SEM are tensor-products of $N$th-order Lagrange interpolants on the Gauss
Lobatto Legendre (GLL) nodal points inside each element.  In our SEM-based
formulation, we solve the unsteady INSE (\ref{eq:nsemomentum},\ref{eq:nsemass})
in the velocity-pressure form using semi-implicit BDF$k$/EXT$k$ timestepping in
which the time derivative is approximated by a $k$th-order backward difference
formula (BDF$k$), the nonlinear terms (and any other forcing) are treated with
$k$th-order extrapolation (EXT$k$), and the viscous and pressure terms are
treated implicitly.  This approach leads to a \emph{linear unsteady Stokes
problem} to be solved at each timestep, which is split into independent viscous
and pressure (Poisson) updates \cite{tomboulides1997}.  The SEM formulation for
the incompressible Navier-Stokes equations in a single conforming domain has
been discussed in detail by Deville, Fischer, and Mund \cite{dfm02},
and we provide a summary of the formulation in \cite{mittal2019nonconforming}.

\begin{figure}[t!] \begin{center}
$\begin{array}{cc}
\includegraphics[height=43mm]{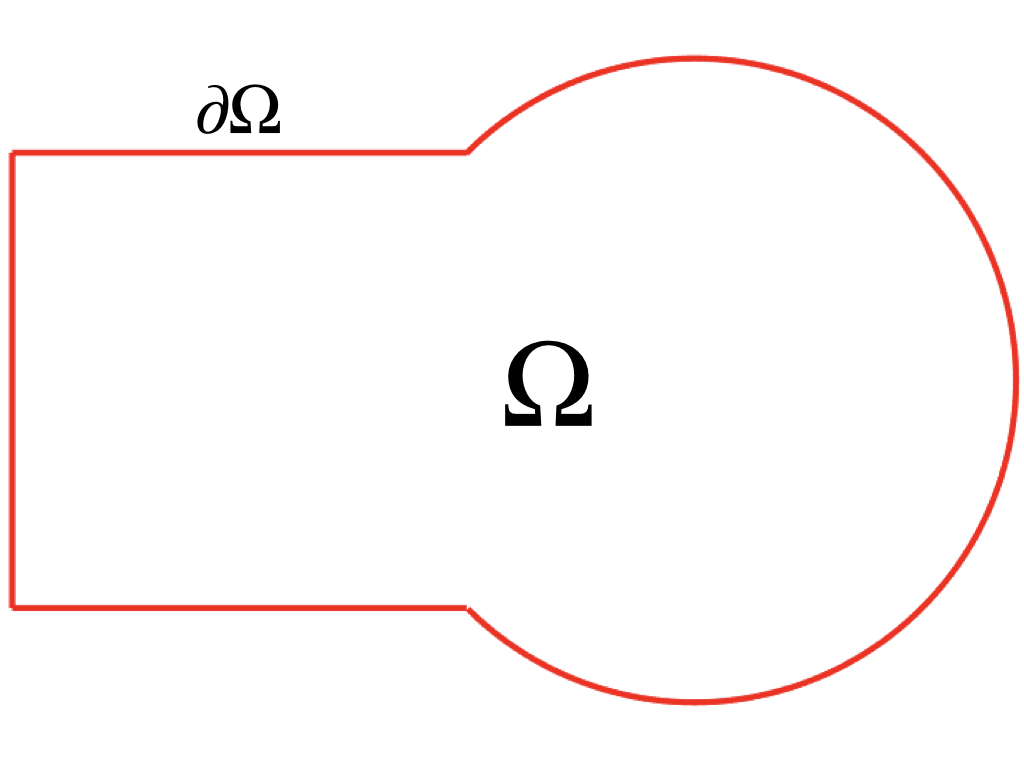} & \hspace{2mm}
\includegraphics[height=45mm]{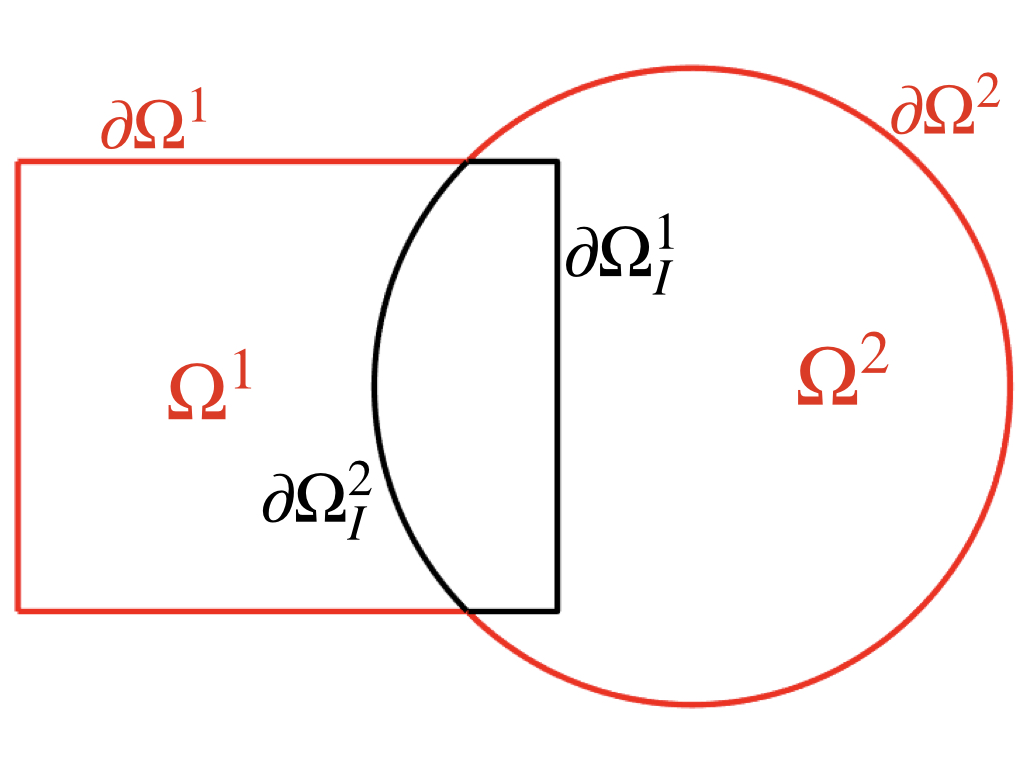}
\end{array}$
\end{center}
\vspace{-7mm}
\caption{(left to right) (a) Composite domain $\Omega$ (b) modeled by overlapping
rectangular ($\Omega^1$) and circular ($\Omega^2$) subdomains. $\dO^{s}_{I}$
denotes the segment of the subdomain boundary $\dO^s$ that is interior to
another subdomain $\Omega^r.$}
\label{fig:schwarz}
\end{figure}

The SEM was extended to the Schwarz-SEM framework for solving the INSE in
overlapping subdomains by Merrill \cite{merrill2016}. In the Schwarz-SEM
framework, the solution to the INSE is advanced in time using the same approach
as that for monodomain SEM, with information exchange at each timestep to
ensure solution consistency in the subdomain overlap region.  Figure
\ref{fig:schwarz} shows an example of the domain $\Omega$ modeled using $S=2$
overlapping subsets. As we can see, overlapping subdomains introduces
``interdomain boundaries'', namely the segments of the subdomain boundary
$\dO^s$ that are interior to another subdomain, that require boundary data to
be interpolated from the corresponding overlapping subdomain.  The interdomain
boundaries are $\dO^1_{I}:=\dO^{1} \subset \Omega^2$ and $\dO^2_{I}:=\dO^{2}
\subset \Omega^1$ for this example and are highlighted in Fig.
\ref{fig:schwarz}(b).

Since overlapping grids rely on interpolation for interdomain boundary data,
a robust and accurate interpolation operator is crucial for OS-based methods.
In the Schwarz-SEM framework, interdomain boundary data interpolation is
effected via \fpt, a scalable high-order interpolation library \cite{mittal2019nonconforming,findpts}.
\fpt\ provides two key functionalities. First, for a given
set of interdomain boundary points that are tagged with the associated subdomain
number $s$, $\bx^* = (\bx_1^*, \bx_2^* \dots \bx_b^*)^s$,
\fpt\ determines the computational coordinates of each point. These computational
coordinates ($\bq={r,e,\bxi,p}$) for each point specify the subdomain number $r$
that it overlaps, the element number ($e \in \Omega^r$) in which the
point was found, and
the corresponding reference-space coordinates ($\bxi = (\xi,\eta,\zeta)$)) inside that
element. Since a mesh could be partitioned on to many MPI ranks, \fpt\
also specifies the MPI rank $p$ on which the donor element is located.
Second,
for a given set of computational coordinates, \fpt\ can interpolate any scalar
function defined on the spectral element mesh.

For static domains, the computational coordinate search only needs to be done
at the beginning of a calculation. For moving domains however, this
identification must be done everytime the location of
$\dO_I^s$ changes with respect to $\Omega^r$. In our numerical calculations
for studying flow past a rotating particle, we use \fpt\ to do computational
coordinate search at each time-step and interpolate the
interdomain boundary data between the rotating inner mesh and the static
background mesh.

With a mechanism for interdomain boundary data interpolation, the solution to
the INSE is advanced in time using a predictor-corrector approach. For the
solution at discrete time $t^n$, since the solution is only known up to time
$t^{n-1}$, the predictor step solves the unsteady Stokes problem with
$m$th-order temporal extrapolation of the spatially-interpolated interdomain
boundary data to maintain the temporal accuracy of the underlying BDF$k$/EXT$k$
timestepper (typically $m=k$). While this approach ensures high-order temporal
accuracy of $\bu(\bx,t^n)$ and $p(\bx,t^n)$, it is not stable and requires $Q$
corrector iterations\footnote{Numerical experiments show that for third-order
temporal accuracy (i.e., $m=3$), $Q=1-3$ is typically sufficient to ensure a
stable solution.} \cite{peet2012}.  Prior to each corrector iteration, the
interdomain boundary data is interpolated between overlapping subdomains, and
the velocity and pressure solution is updated using the unsteady Stokes
problem.

The Schwarz-SEM framework introduced in \cite{merrill2016} was originally
developed for two static overlapping grids. Recently, this framework has been improved
to support moving subdomains using the ALE formulation \cite{merrill2019moving},
along with support for an arbitrary number of overlapping grids \cite{mittal2019nonconforming}.
The Schwarz-SEM framework has demonstrated that it maintains the spatial and temporal convergence of the underlying SEM solver for an arbitrary number of static and moving overlapping grids, and
accurately models highly turbulent flow in complex domains \cite{mittal2019highly}.
Additionally, the Schwarz-SEM framework has shown that it can significantly
reduce the computational cost in comparison to a single conforming grid-based calculation.
Since overlapping grids support nonconforming overlap, each grid can be constructed
based on the physics of the flow in the region that they cover. This flexibility
in grid construction has been used to reduce the total element count by as much as
50\% (e.g., see Section 9.5 in \cite{mittal2019highly}).

In cases similar to thermal plumes or jets where the Courant-Friedrichs-Lewy (CFL) number
varies significantly throughout the domain, the Schwarz-SEM framework has been
recently improved to enable
multirate timestepping (MTS) for INSE such that each subdomain can use a timestep
size based on its local CFL \cite{mittal2020multirate}. Multirate timestepping
is not possible in the case of a single conforming grid because
of the tight coupling implied by the incompressibility constraint, which leads to
use of a constant timestep size throughout the domain.
With MTS, different timestep sizes can be used in different subdomains, which
reduces the computational resources required for the subdomain with slower time-scales
because they have to take fewer timesteps in comparison to the subdomain with faster time-scales.
Note that all the results presented here use the singlerate timestepping based formulation,
and we will consider MTS for future calculations with hundreds of rotating particles.

\section{Results}

\subsection{Rotating Spherical Particles}
As a first step towards validating the Schwarz-SEM framework for moving meshes,
we model a rotating sphere at $Re=300$.  Following Dobson et al. \cite{dobson2014flow}, the particle
rotation is set normal to the flow direction, and the nondimensional rotation
rate is set to $\Omega^*=1$. All the results presented in this paper
were obtained with the Schwarz-SEM framework where we used $m=3$ for third-order
temporal accuracy of the solution and $Q=3$ corrector iterations
at each timestep for stability. The results were spatially converged at $N=7$.

\begin{figure}[t!] \begin{center}
$\begin{array}{c}
\includegraphics[width=75mm]{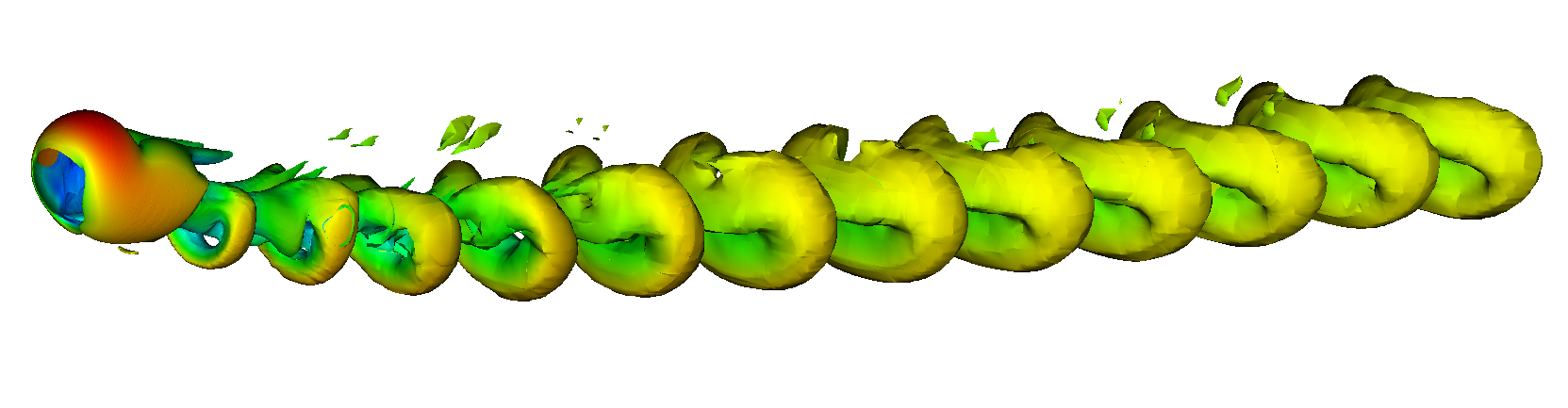} \\
\includegraphics[width=75mm]{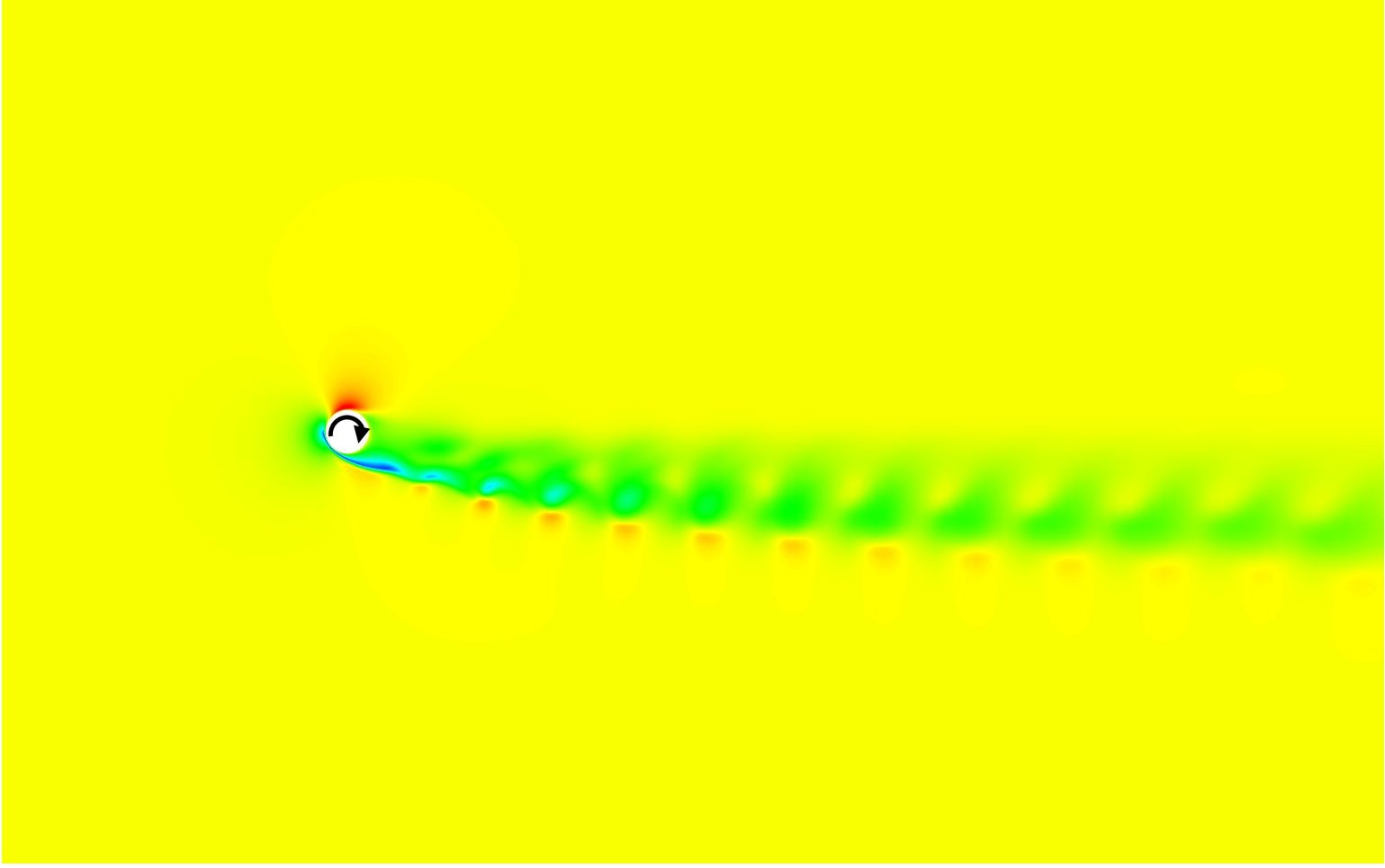} \\
\includegraphics[width=55mm]{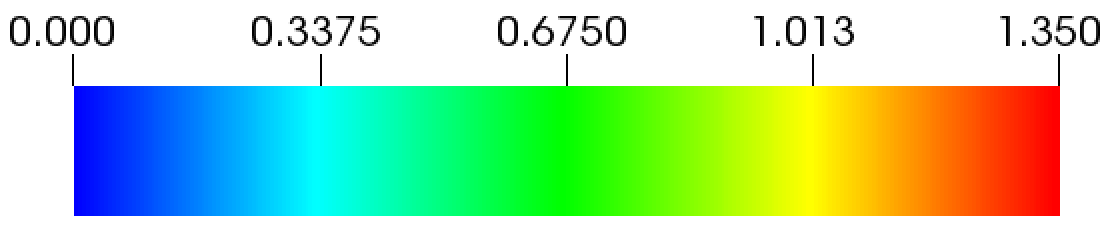}
\end{array}$
\end{center}
\vspace{-5mm}
\caption{(top) Isosurfaces of $\lambda_{2}$ colored by velocity magnitude, and
(bottom) velocity magnitude contours for flow over the rotating sphere.}
\label{fig:rotsph1_vel}
\end{figure}

Figure \ref{fig:rotsph1_vel} shows isosurface of $\lambda_2$ colored by velocity magnitude (top), and the velocity magnitude contour for the the rotating sphere. The isosurface of $\lambda_2$ shows the topology and geometry of the vortex core \cite{lambda2},
illustrating the coherent vortical structure generated behind the rotating particle.
Figure \ref{fig:rotsph1_vel} shows vortex shedding in the wake of the sphere, with the
flow separating around the point of least relative velocity between the sphere
and the background flow. Note that the rotation of the sphere makes the vortex shedding
asymmetrical, when compared against the non-rotating case.

\begin{figure}[t!] \begin{center}
$\begin{array}{cc}
\includegraphics[width=45mm]{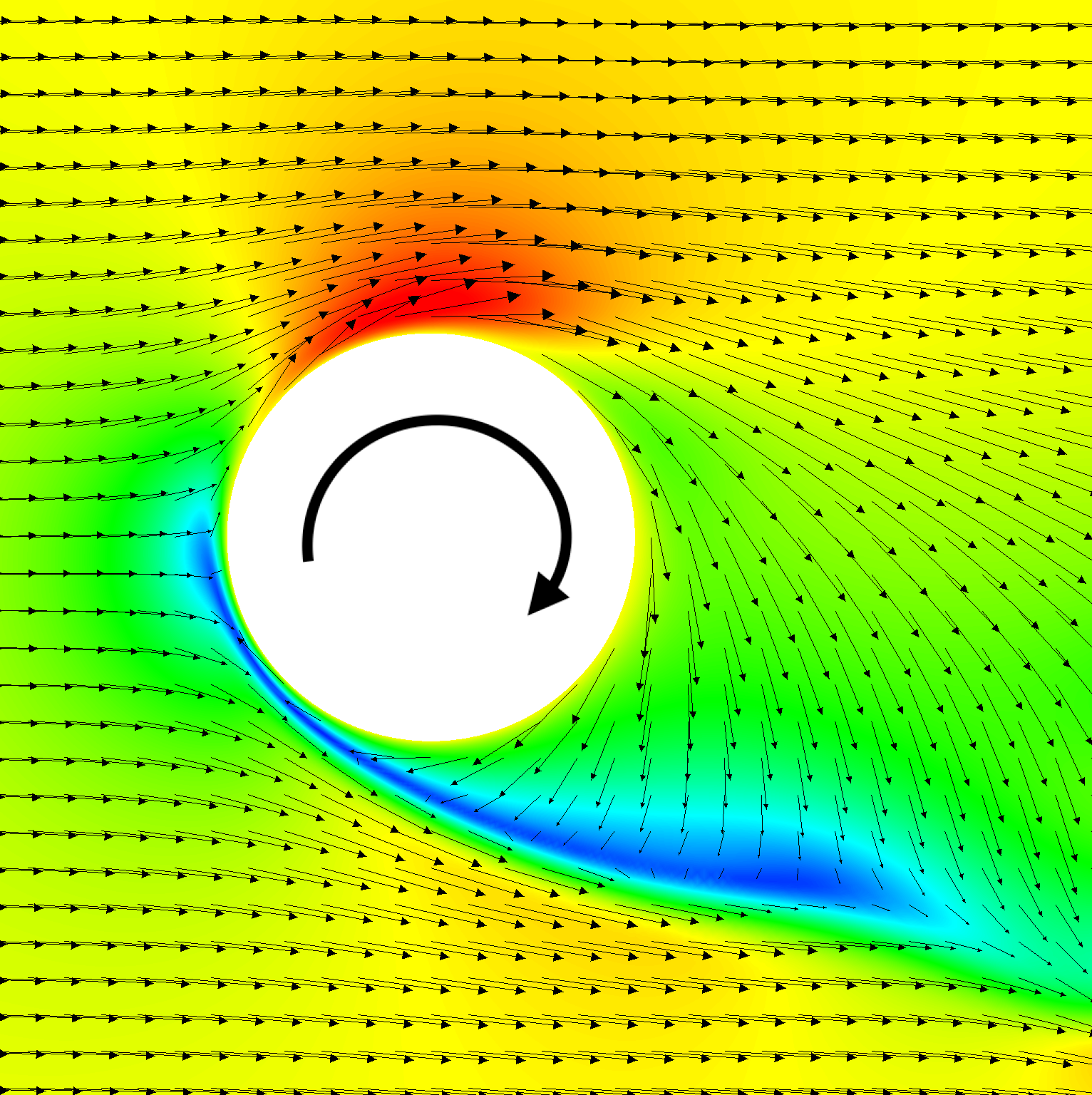} &
\includegraphics[width=45mm]{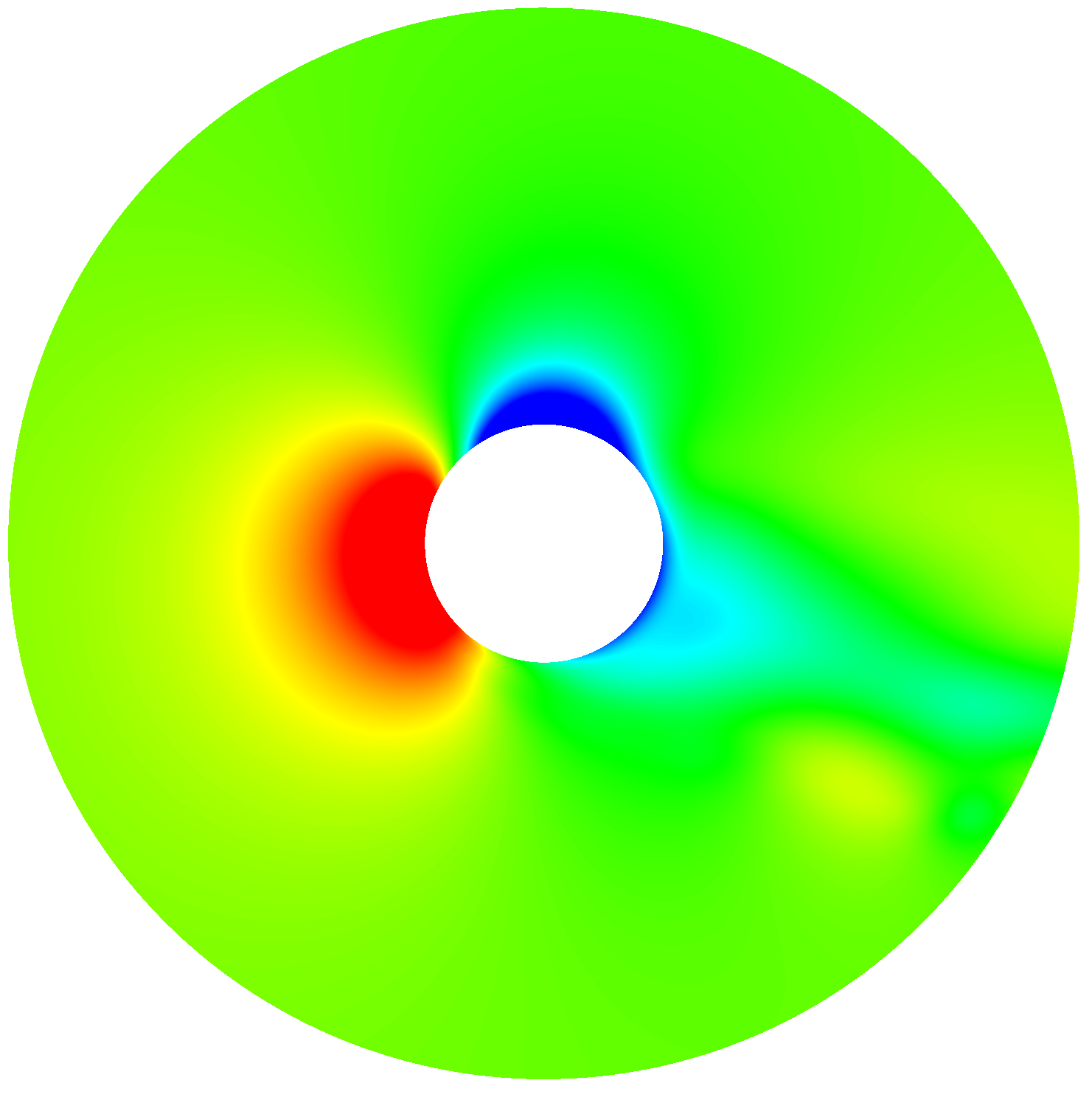} \\
\multicolumn{2}{c}{\includegraphics[width=100mm]{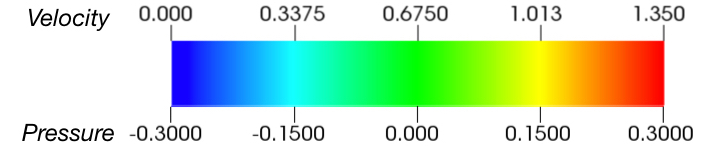}}
\end{array}$
\end{center}
\vspace{-7mm}
\caption{Slice view of the (left) velocity vector field and (right) pressure
near the rotating sphere.}
\label{fig:rotsph1_vel2}
\end{figure}

Figure \ref{fig:rotsph1_vel2} shows flow around the rotating
sphere, on a y-z plane passing thorough the center of the sphere.
We can observe in the velocity vector plot in Fig. \ref{fig:rotsph1_vel2} that the rotating sphere pulls fluid from the leeward
side to the windward side\footnote{The windward side refers to the side of the
sphere which is facing the background flow, and leeward side refers to the side that
does not directly interact with the background flow.}, which is similar to the
phenomena of added-mass where fast-moving/accelerating particles are known to
carry additional mass of fluid around them \cite{schwarzkopf2011}.
Consequently, a shear layer forms between the opposing flows of the fluid
pulled by the rotating sphere and the background flow along the
$z$-axis.  With the shear layer instability growing behind the sphere, we see
vortex shedding about one sphere diameter downstream of the sphere.  This
vortex shedding is also apparent in Fig. \ref{fig:rotsph1_vel}.  Due to the
fluid being pulled from the leeward to the windward side and its interaction
with the background flow, a high pressure region forms at the bottom of the
sphere. This relatively high pressure region at the bottom along with a
relatively low pressure region on the top of the sphere leads to lift force
(along the $y$ direction) on the sphere. Similarly, a high pressure region on
the windward side of the sphere, due to formation of a stagnation point, and a low pressure
region on the leeward side, due to the rotation of the sphere, results in the
\textit{form} drag (along $z$ direction) acting on the particle. The flow structures that we see in Fig.
\ref{fig:rotsph1_vel} and Fig.  \ref{fig:rotsph1_vel2} are similar to those
reported by Dobson \cite{dobson2014flow} and others
\cite{kim2009laminar,poon2010laminar}.

Table \ref{table:rotsphdrag} lists the lift coefficient ($C_{Ly}$), drag
coefficient ($C_{Dz}$) and Strouhal number ($St$) determined by the Schwarz-SEM simulations for the rotating sphere at $Re=300$ with $\Omega^*=1$ and 2. The results are
compared with Dobson et al. (and other literature for the case of
$\Omega^*=1$). The drag coefficient was computed as $C_{Dz} = 2F_z/(\rho
U_{\infty}^2A_p)$, where $F_z$ is the force on the particle in the streamwise
direction and $A_p=\pi D^2/4$ is the projected area of the sphere. The lift
coefficient is computed similarly using the lift force. Table
\ref{table:rotsphdrag} shows that our results match the data from Dobson et al.
to within 1.5\%.

\begin{table}[t!]
\begin{center}
 \begin{tabular}{|p{45mm} | c | c | c|}
 \hline
  & $C_{Ly}$ & $C_{Dz}$ & $St$ \\
 \hline
 \multicolumn{4}{|c|}{$\Omega^*=1$}  \\
 \hline
 Schwarz-SEM     & 0.613 & 0.961 & 0.426 \\
 \hline
 Dobson et al. \cite{dobson2014flow}   & 0.610 & 0.961 & 0.423 \\
 \hline
 Kim et al. \cite{kim2009laminar}   & 0.596 & 0.931 & 0.424 \\
 \hline
 Poon et al. \cite{poon2010laminar}   & 0.605 & 0.964 & 0.427 \\
 \hline
  \multicolumn{4}{|c|}{$\Omega^*=2$}  \\
  \hline
  Schwarz-SEM     & 0.589 & 1.019 & 0.243 \\
  \hline
  Dobson et al.   & 0.582 & 1.012 & 0.240 \\
  \hline
\end{tabular}
\caption{Comparison of lift coefficient ($C_{Ly}$), drag coefficient ($C_{Dz}$), and Strouhal number
($St$) for the Schwarz-SEM calculations, with Dobson et al. and others.}
\label{table:rotsphdrag}
\end{center}
\end{table}

\subsection{Rotating Ellipsoidal Particles}
With the Schwarz-SEM framework validated using a rotating sphere, we now consider
two different ellipsoidal particles. We keep
$a_x=a_z=0.5D$ unchanged from the sphere in the previous section, and modify
$a_y=0.75D$ for one particle and $a_y=0.25D$ for the other.
Thus, the ratios of $y$ and $z$ axis for the two particles are $1.5$ and $0.5$.
As before, the axis
of rotation is $x$, and the background flow is in the positive $z$ direction.
The computational meshes for these nonspherical (ellipsoid) particles have been generated by
morphing the inner mesh used for the spherical particle. This morphing of mesh
is straightforward to effect since the surface of the particle is described as
$x^2/a_x^2 + y^2/a_y^2 + z^2/a_z^2 = 1$.
The morphed mesh was smoothed to reduce the pressure iterations of the pressure-Poisson solver \cite{mittal2019mesh}.

\begin{figure}[t!] \begin{center}
		$\begin{array}{c}
		\includegraphics[height=80mm]{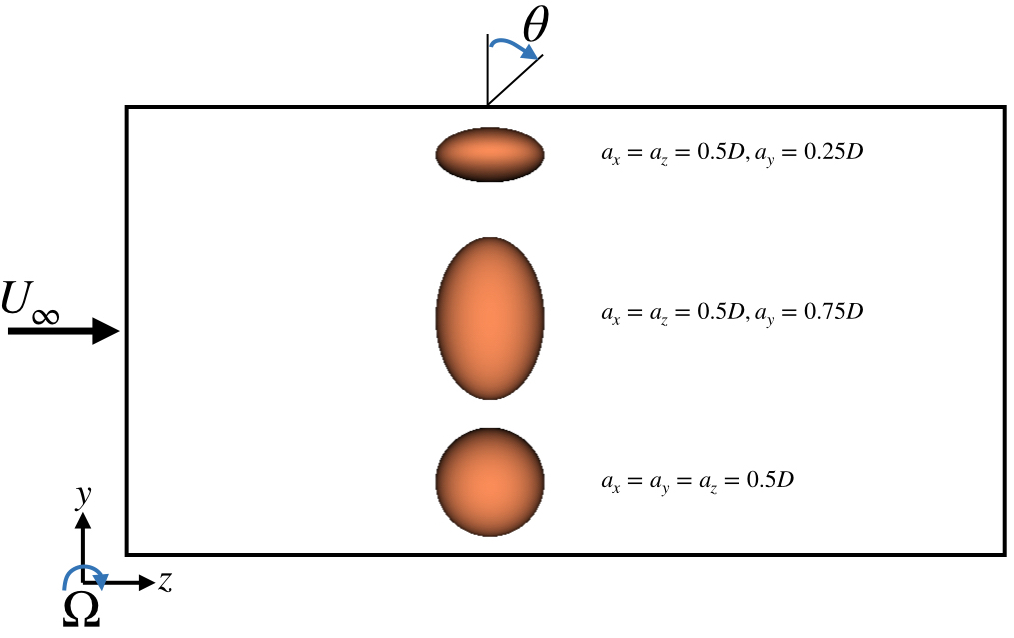}
		\end{array}$
	\end{center}
	\caption{Schematic showing the direction of the rotation of particle with
		respect to the flow for different particles, and the shape of three  particles
		considered in the current study.}
	\label{fig:spheroidschematic}
\end{figure}

Figure \ref{fig:spheroidschematic}
contrasts the two ellipsoids with the spherical geometry of the preceding
example.
The position of the particles shown in Fig.
\ref{fig:spheroidschematic} corresponds to $\theta=0^\deg$, and $\theta$
increases from $0^\deg$ to $360^\deg$ as the particle rotates clockwise around
the $x$-axis. Initially, the calculations for the rotating ellipsoid (and sphere) were conducted with a much coarser mesh resolution. These coarser domains had a total of about $E=10,000$ spectral elements and used $N=7$ (total grid-points is $\approx n = EN^3 = 3.43$ million). To check grid independence, the resolution was incrementally raised to about $E=80,000$ ($n = 27$ million) spectral-elements. Between $E=10,000$ and $E=80,000$, increasing the resolution did not substantially change the phase-averaged drag and lift; though the time-series of instantaneous drag and lift became smoother with increase in resolution.

\begin{figure}[t!] \begin{center}
                $\begin{array}{c}
                \includegraphics[width=80mm]{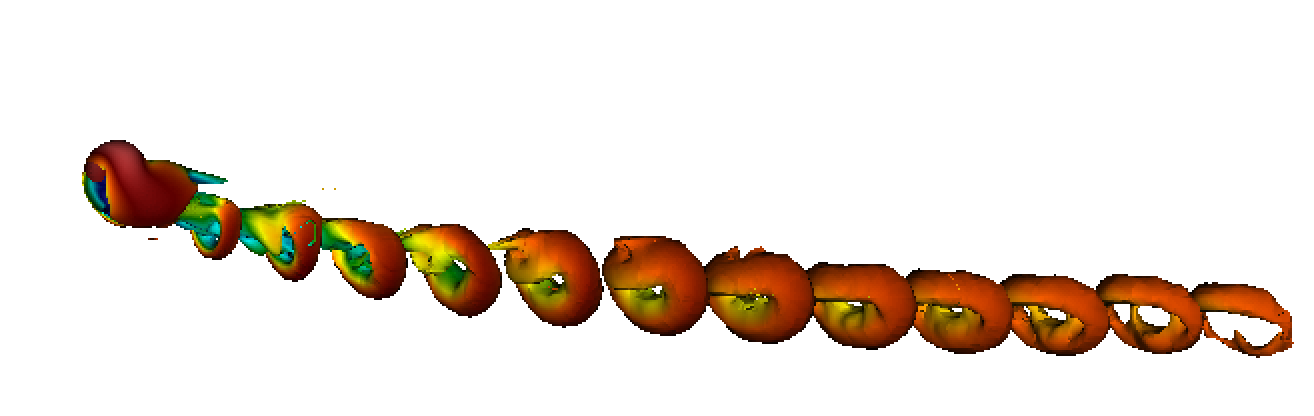} \\
                \textrm{(a) } a_x=a_y=a_z=0.5D \\
                \includegraphics[width=80mm]{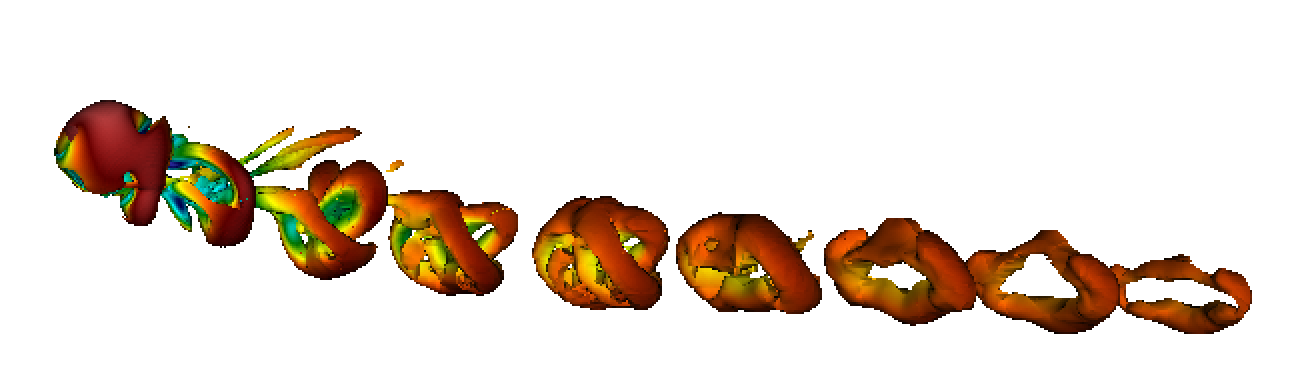} \\
                \textrm{(b) } a_x=a_z=0.5D, a_y = 0.75D \\
                \includegraphics[width=80mm]{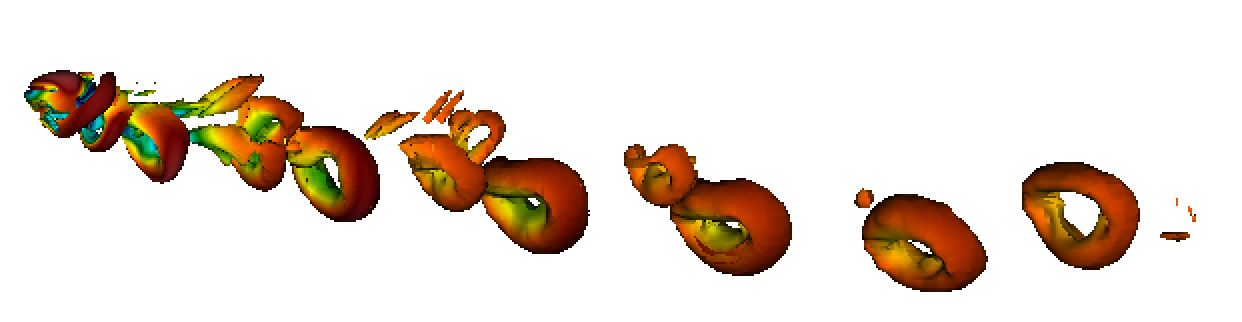} \\
                \textrm{(c) } a_x=a_z=0.5D, a_y = 0.25D\\
                \includegraphics[width=40mm]{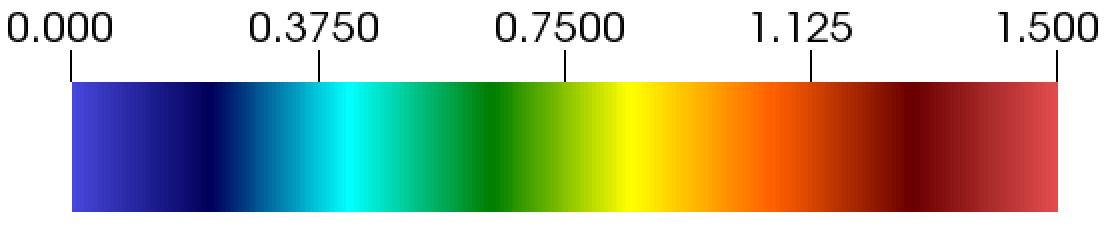}
                \end{array}$
        \end{center}
        \caption{Comparison of $\lambda_2$  iso-surfaces \cite{lambda2} (illustrating the topology and shape of the vortex core), colored by velocity magnitude, for the three rotating particles considered in this study.}
        \label{fig:rotsph_vort}
\end{figure}

Figure \ref{fig:rotsph_vort} shows $\lambda_2$ \cite{lambda2} for the three
particles simulated in the current study. It is apparent from Fig.  \ref{fig:rotsph_vort} that
the structure of the vortices being shed behind the ellipsoids is different
from the structures that we had observed for the sphere. The primary reason
behind this difference is that unlike in the flow over the rotating sphere, a
steady shear layer never develops behind the ellipsoid. The flow keeps
attaching and separating as the ellipsoid rotates, which leads to multiple
vortices being shed in its wake. The primary vortex shedding mechanism is the
interaction of the mean background flow with the ellipsoid, which we had
observed for the spherical particle as well. For the ellipsoids, there is also
a secondary mechanism due to the asymmetry in shape. This mechanism will be
clear through the discussion in the next section.

Since the shape of the particle has changed, it is essential to consider what
area to use for computing the drag (and lift) coefficient, $C_D = 2F/(\rho
U_{\infty}^2 A_p)$. For the case of drag in the streamwise direction, the
choice of the frontal area is straightforward.  However, for the lift
coefficient (normal to the direction of the flow - $y$ in Fig.
\ref{fig:spheroidschematic}), it is not clear whether the projected area normal
to the direction of the flow is the best choice. Using the frontal area of the
sphere (constant at $\pi D^2/4$) indicates that decreasing $a_y$ increases the
lift coefficient. However, using the actual frontal area of the ellipsoids
(time-varying due to rotation) indicates that decreasing $a_y$ decreases the
lift coefficient. Thus, instead of comparing the drag and lift coefficients, we
compare the nondimensionalized drag and lift forces to avoid any confusion and
inconsistencies.

\begin{figure}[b!] \begin{center}
     $\begin{array}{ccc}
      \hspace{-30mm}
      \includegraphics[width=65mm]{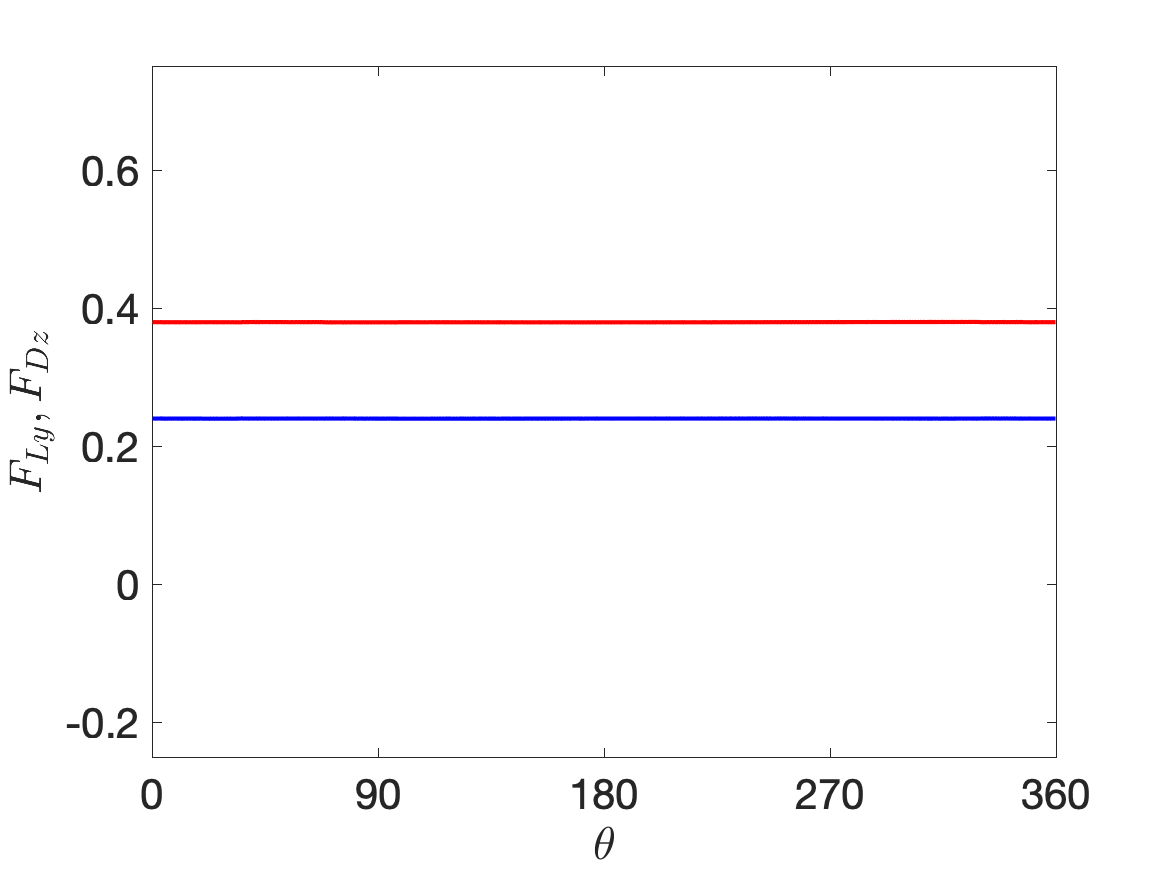} &
      \hspace{-05mm}
      \includegraphics[width=65mm]{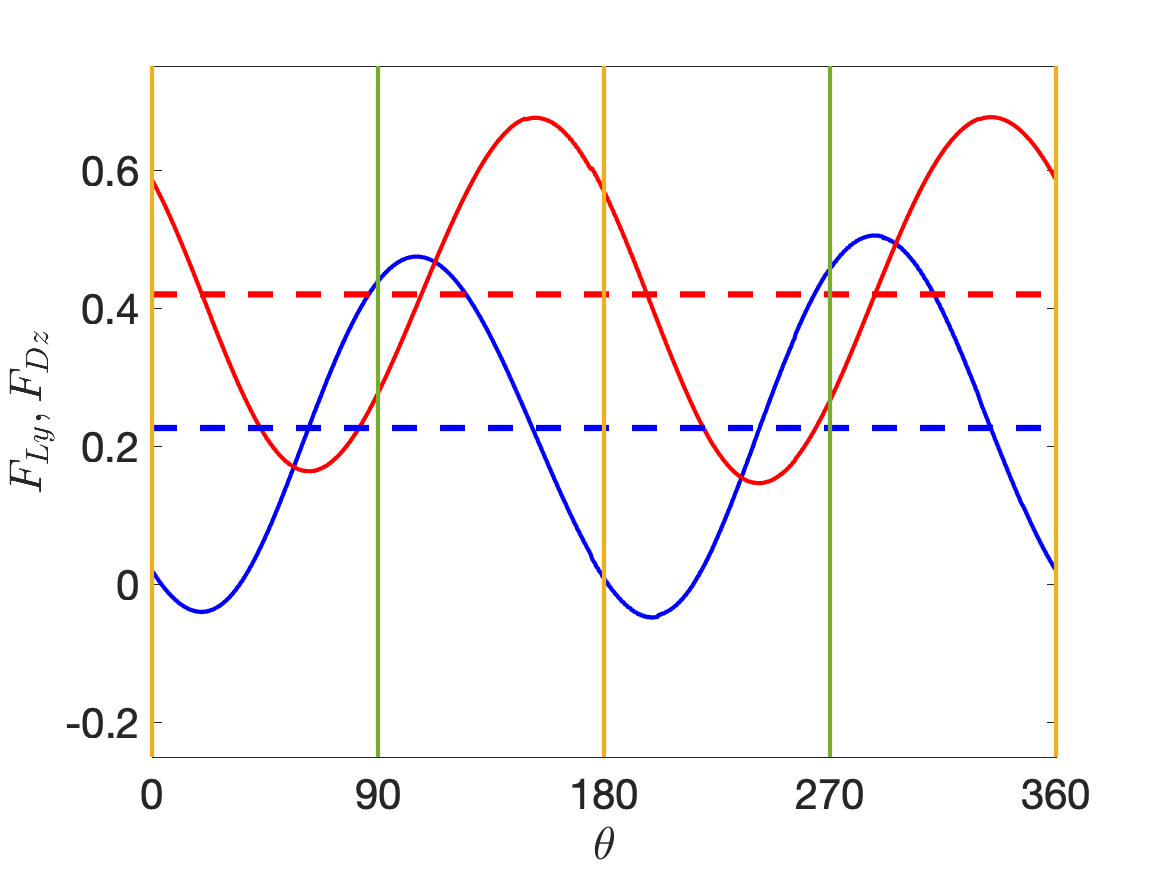} &
      \hspace{-05mm}
      \includegraphics[width=65mm]{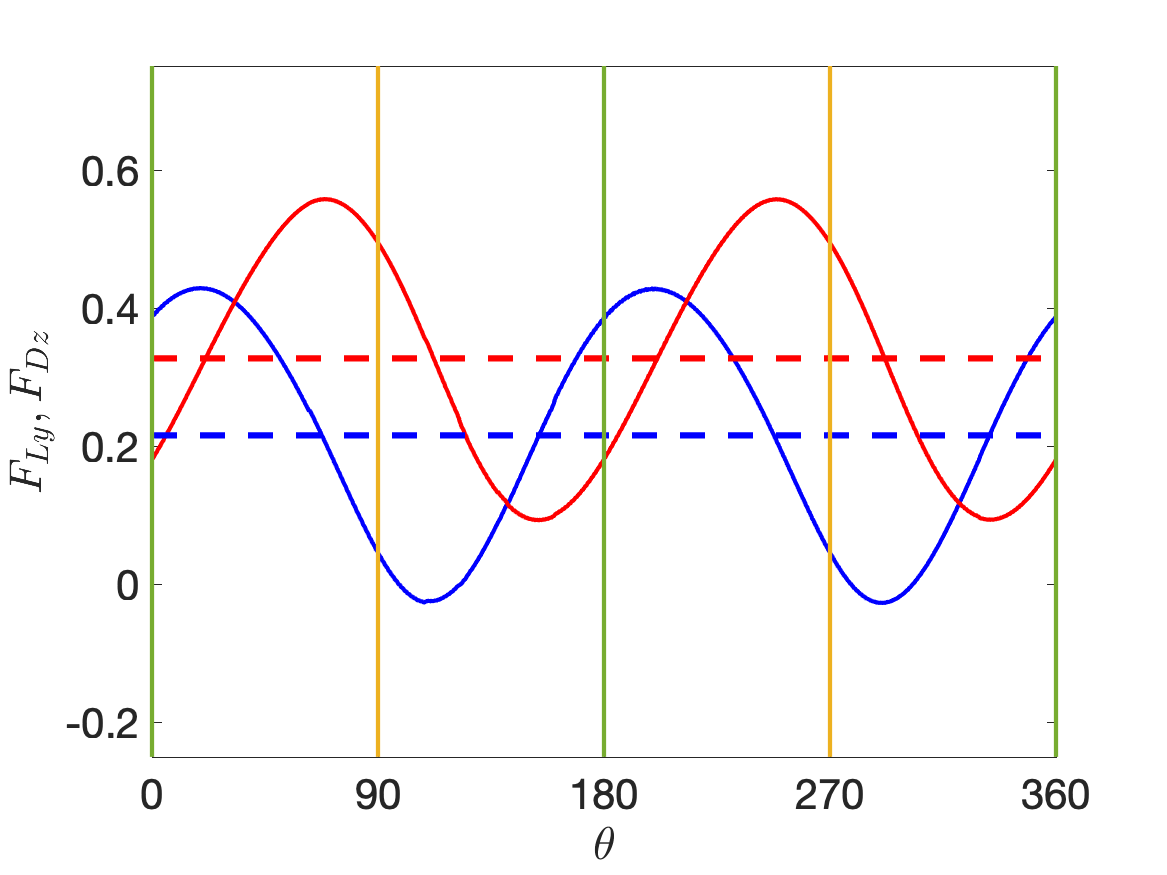} \\
      \multicolumn{3}{c}{\includegraphics[width=50mm]{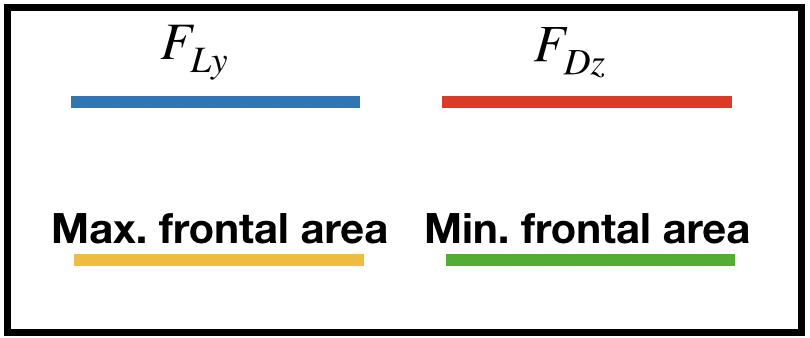}}
     \end{array}$
     \end{center}
\caption{Comparison of the phase-averaged drag and lift with rotation angle ($\theta$) for the
(left) spherical particle; and ellipsoids with (center) $a_y=0.75D$ and (right)
$a_y=0.25D$. The dashed lines in the (center) and (right) plot indicate the
mean values for the drag and lift on each ellipsoid. The golden- and
green-colored vertical lines indicate the angles at which the frontal area of the
particle is maximum and minimum, respectively.}
\label{fig:rotsph_e1draglift}
\end{figure}

Figure \ref{fig:rotsph_e1draglift} compares the phase-averaged drag and lift on the three
rotating particles with $\theta$. The drag and lift forces on the sphere are
almost constant in time because the flow is steady near the sphere surface. In
contrast to the sphere, we observe that the drag and lift vary in time for the
ellipsoids, and there is a phase-shift of $90^\deg$ for the drag and lift
time-series for the two ellipsoids. This phase-shift is expected because the
major principal axis of the particle with $a_y=0.75D$ is the minor principal
axis for the particle with $a_y=0.25D$. In Fig.  \ref{fig:rotsph_e1draglift},
we have also indicated $\theta$ at which the frontal area of the particles
is maximum and minimum, using vertical golden- and green-colored lines, respectively.
An interesting observation from these results is that there is a phase difference
between the $\theta$ of maximum (or minimum) frontal area and the $\theta$ at
which the drag/lift is maximum (or minimum).

\begin{table}[b!]
\begin{center}
        \begin{tabular}{|p{60mm} | c | c | c|}
                \hline
                & $F_{Ly}$ [$F_{Ly,min},F_{Ly,max}$] & $F_{Dz}$ [$F_{Dz,min},F_{Dz,max}$] \\
                \hline
      $a_x=a_y=a_z=0.5D$     & 0.240 & 0.379 \\
                \hline
      $a_x=a_z=0.5D, a_y = 0.75D$  & 0.226 [-0.048,0.505] & 0.419 [0.146,0.676] \\
                \hline
      $a_x=a_z=0.5D, a_y = 0.25D$  & 0.216 [-0.028,0.429] & 0.327 [0.092,0.558] \\
                \hline
        \end{tabular}
        \caption{Comparison of lift ($F_{Ly}$) and drag ($F_{Dz}$) for the Schwarz-SEM
                calculations.}
        \label{table:rotpartdrag}
\end{center}
\end{table}

Table \ref{table:rotpartdrag} lists the mean drag and lift forces on each
particle for $Re=300$ at $\Omega^*=1$ along with the maximum and minimum drag
and lift forces.  As we can see, increasing $a_y$ increases the mean drag on
the particle, and similarly decreasing $a_y$ decreases the mean drag.
This increase/decrease is expected due to change in the frontal area
of the particle. Though, the maximum and minimum drag registered for both the ellipsoidal particles is substantially different from the sphere. This clearly shows the significant role a small change in shape can have on the dynamics of the particle.
On the other hand, mean lift acting on the particles ($F_{Ly}$) has decreased
for both the ellipsoids in comparison to the sphere. Despite decrease in the mean lift, the maximum lift force experienced by the ellipsoidal particles is almost twice the lift force on the sphere.

The results presented in Fig. \ref{fig:rotsph_e1draglift} and Table
\ref{table:rotpartdrag} bring forth two key aspects of the flow past a
rotating particle. First, the change in shape of the particle has a
significant impact on the streamwise and transverse forces experienced by the
particle.
Second, the $\theta$ at which the ellipsoids experiences maximum and minimum
streamwise drag ($F_{Dz}$) and transverse lift ($F_{Dy}$), do not coincide with
the $\theta$ at which the frontal area is maximum or minimum.

\section{Discussion}
In this section, we use the ellipsoid with $a_y=0.75D$
to describe the fundamental difference
between the flow over a rotating sphere and a rotating ellipsoid.

\subsection{Effect of rotation and shape on drag and lift}
\begin{figure}[t!] \begin{center}
   $\begin{array}{ccc}
   \includegraphics[width=45mm]{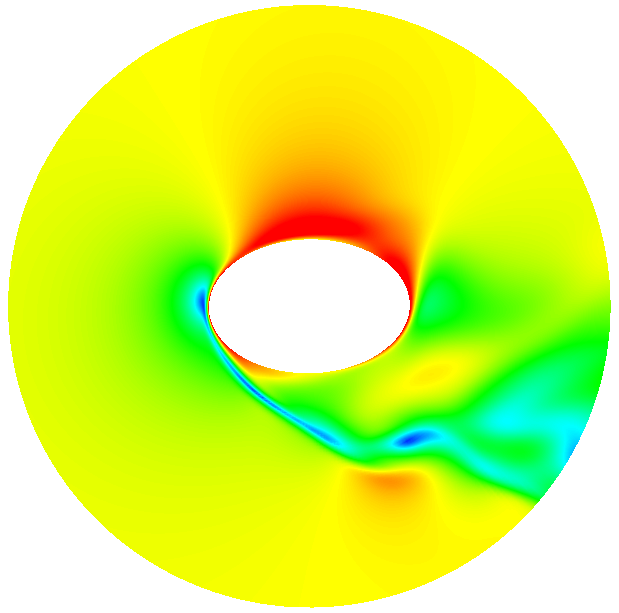} &
   \includegraphics[width=45mm]{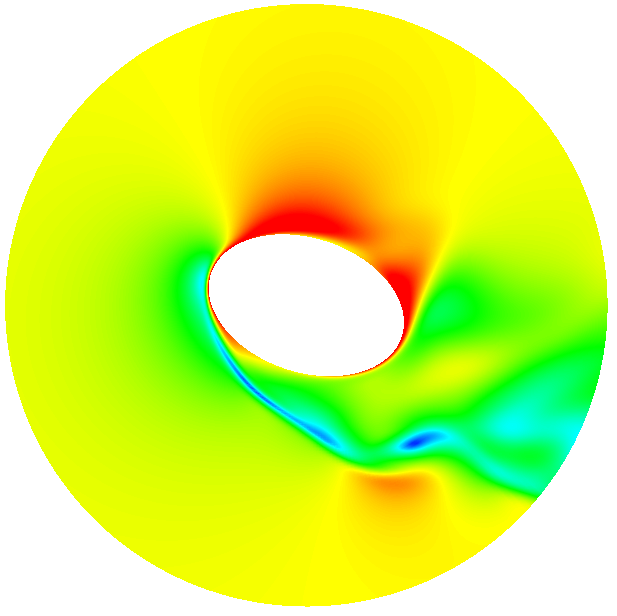} &
   \includegraphics[width=45mm]{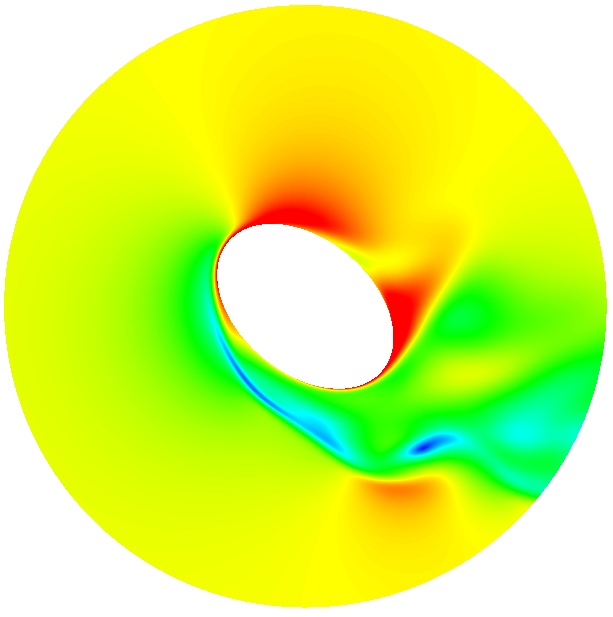} \\
   \textrm{(a) } \theta \approx 90^{\deg} &
   \textrm{(b) } \theta\approx 108^{\deg} \textrm{(maximum lift)} &
   \textrm{(c) } \theta\approx 130^{\deg} \\
   \includegraphics[width=45mm]{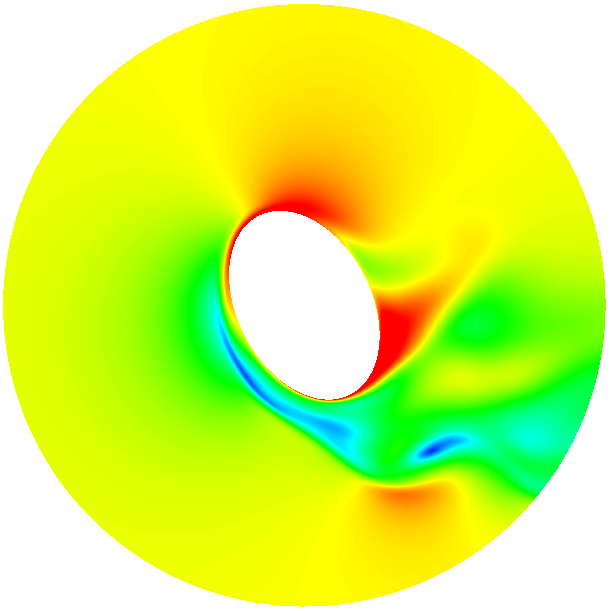} &
   \includegraphics[width=45mm]{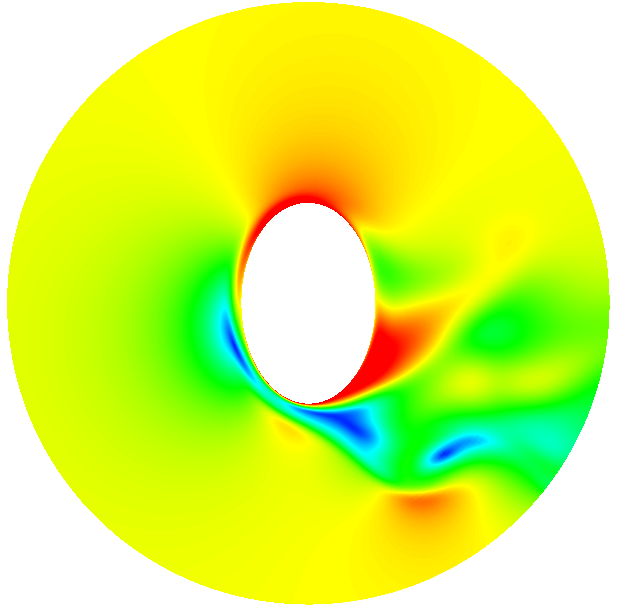} &
   \includegraphics[width=45mm]{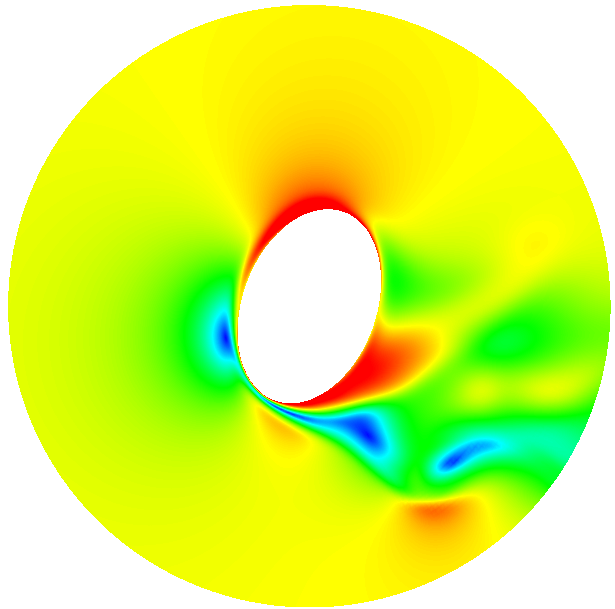} \\
   \textrm{(d) } \theta\approx 153^{\deg} \textrm{(maximum drag)} &
   \textrm{(e) } \theta\approx 180^{\deg} &
   \textrm{(f) } \theta\approx 200^{\deg} \textrm{(minimum lift)} \\
   \includegraphics[width=45mm]{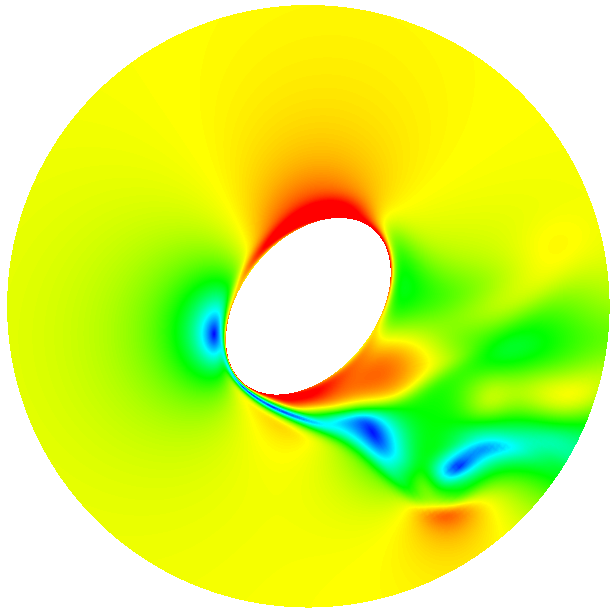} &
   \includegraphics[width=45mm]{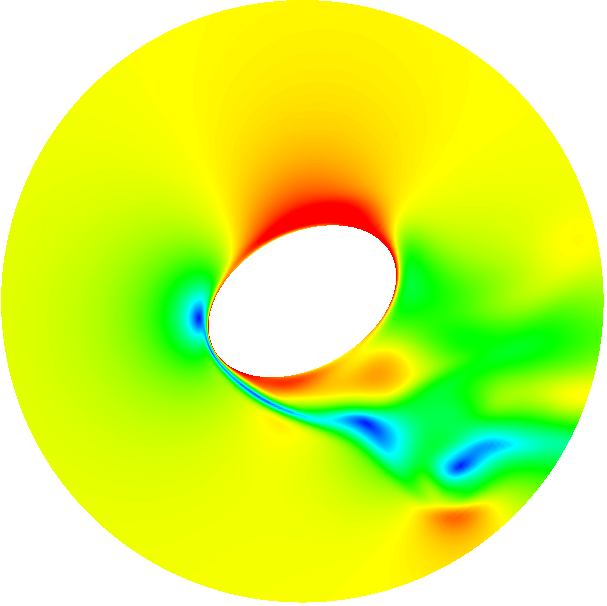} &
   \includegraphics[width=45mm]{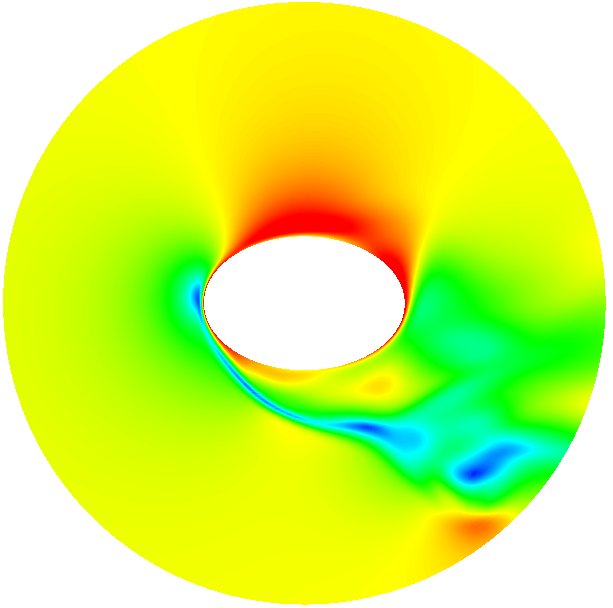} \\
   \textrm{(g) } \theta\approx 220^{\deg} &
   \textrm{(h) } \theta\approx 242^{\deg} \textrm{(minimum drag)} &
   \textrm{(i) } \theta\approx 270^{\deg} \\
   \multicolumn{3}{c}{\includegraphics[width=50mm]{velleg}}
   \end{array}$
 \end{center}
       \hspace{-5mm}
 \caption{Velocity magnitude around the particle on the y-z plane
passing through the center of the ellipsoid ($a_y=0.75D$) at different orientation ($\theta$).}
 \label{fig:e1v_dragmax}
\end{figure}

\begin{figure}[t!] \begin{center}
	$\begin{array}{ccc}
 		\includegraphics[width=45mm]{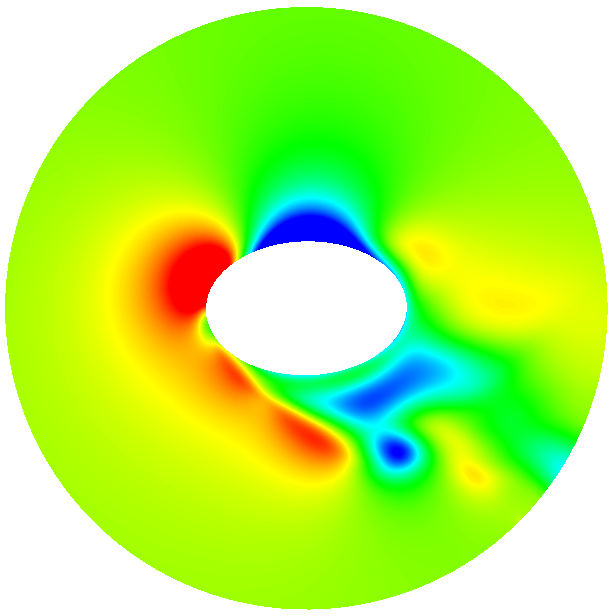} &
 		\includegraphics[width=45mm]{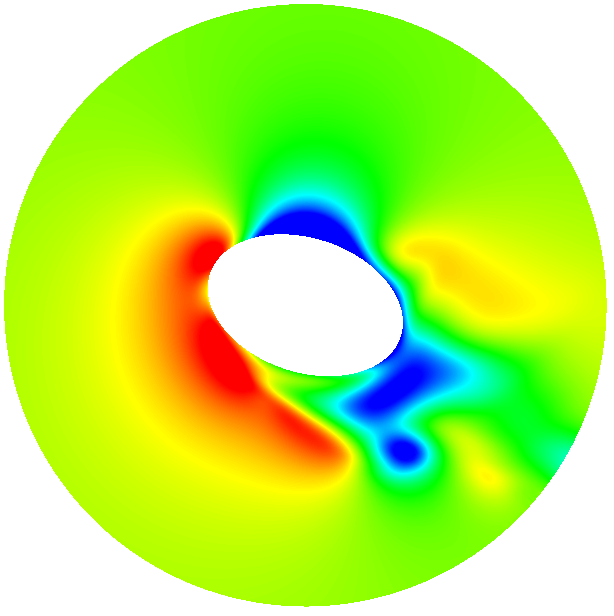} &
 		\includegraphics[width=45mm]{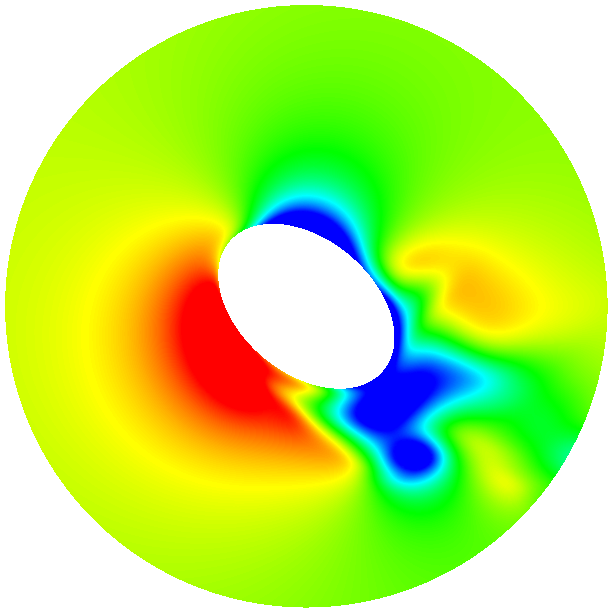} \\
 		\textrm{(a) } \theta \approx 90^{\deg} &
 		\textrm{(b) } \theta\approx 108^{\deg} \textrm{(maximum lift)} &
 		\textrm{(c) } \theta\approx 130^{\deg} \\
 		\includegraphics[width=45mm]{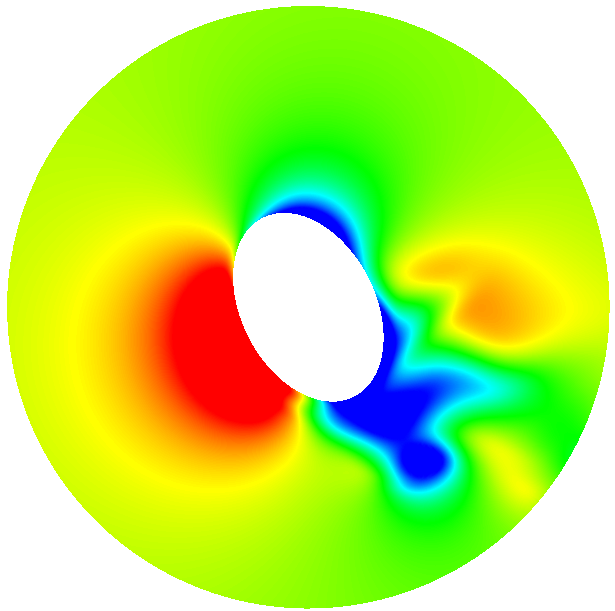} &
 		\includegraphics[width=45mm]{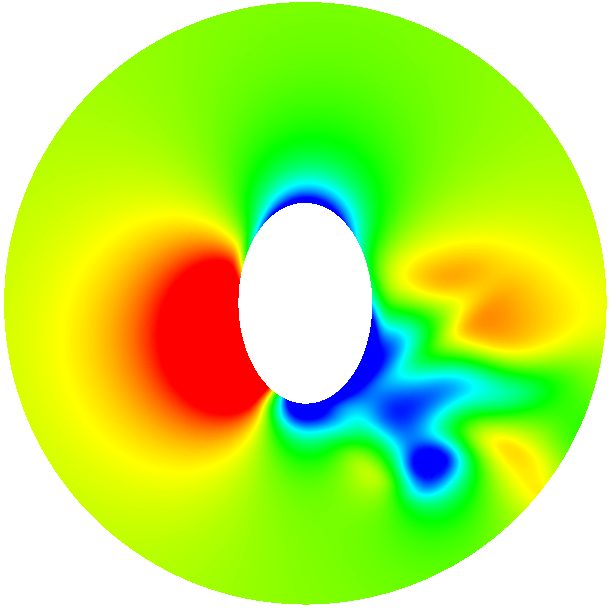} &
 		\includegraphics[width=45mm]{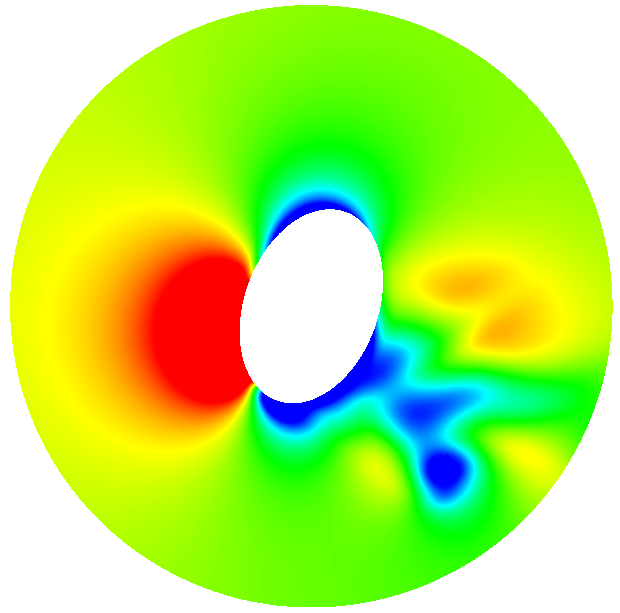} \\
 		\textrm{(d) } \theta\approx 153^{\deg} \textrm{(maximum drag)} &
 		\textrm{(e) } \theta\approx 180^{\deg} &
 		\textrm{(f) } \theta\approx 200^{\deg} \textrm{(minimum lift)} \\
 		\includegraphics[width=45mm]{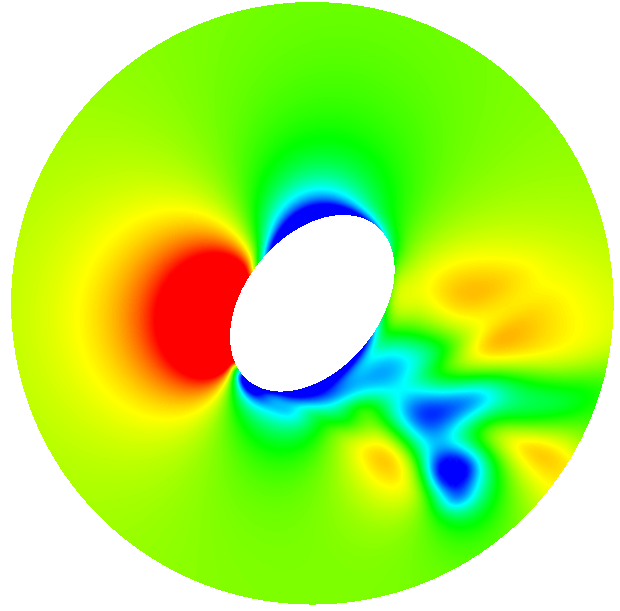} &
 		\includegraphics[width=45mm]{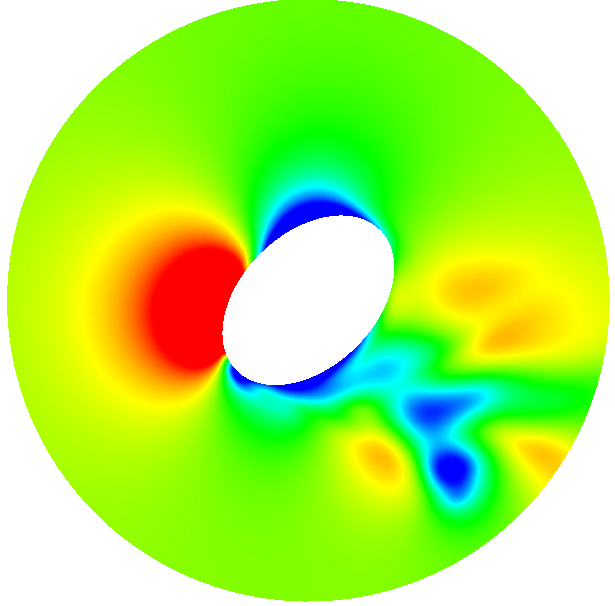} &
 		\includegraphics[width=45mm]{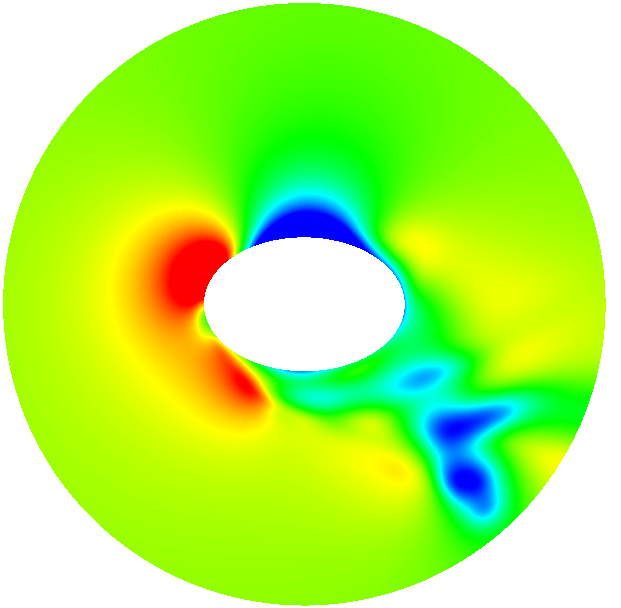} \\
 		\textrm{(g) } \theta\approx 220^{\deg} &
 		\textrm{(h) } \theta\approx 242^{\deg} \textrm{(minimum drag)} &
 		\textrm{(i) } \theta\approx 270^{\deg} \\
 		\multicolumn{3}{c}{\includegraphics[width=50mm]{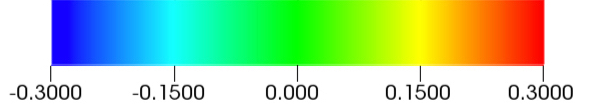}}
 		\end{array}$
 	\end{center}
        \hspace{-10mm}
	\caption{Pressure around the particle on the y-z plane
 passing through the center of the ellipsoid ($a_y=0.75D$) at different orientation ($\theta$).}
 	\label{fig:e1p_dragmax}
 \end{figure}

\begin{figure}[t!] \begin{center}
                $\begin{array}{cc}
                \includegraphics[width=70mm]{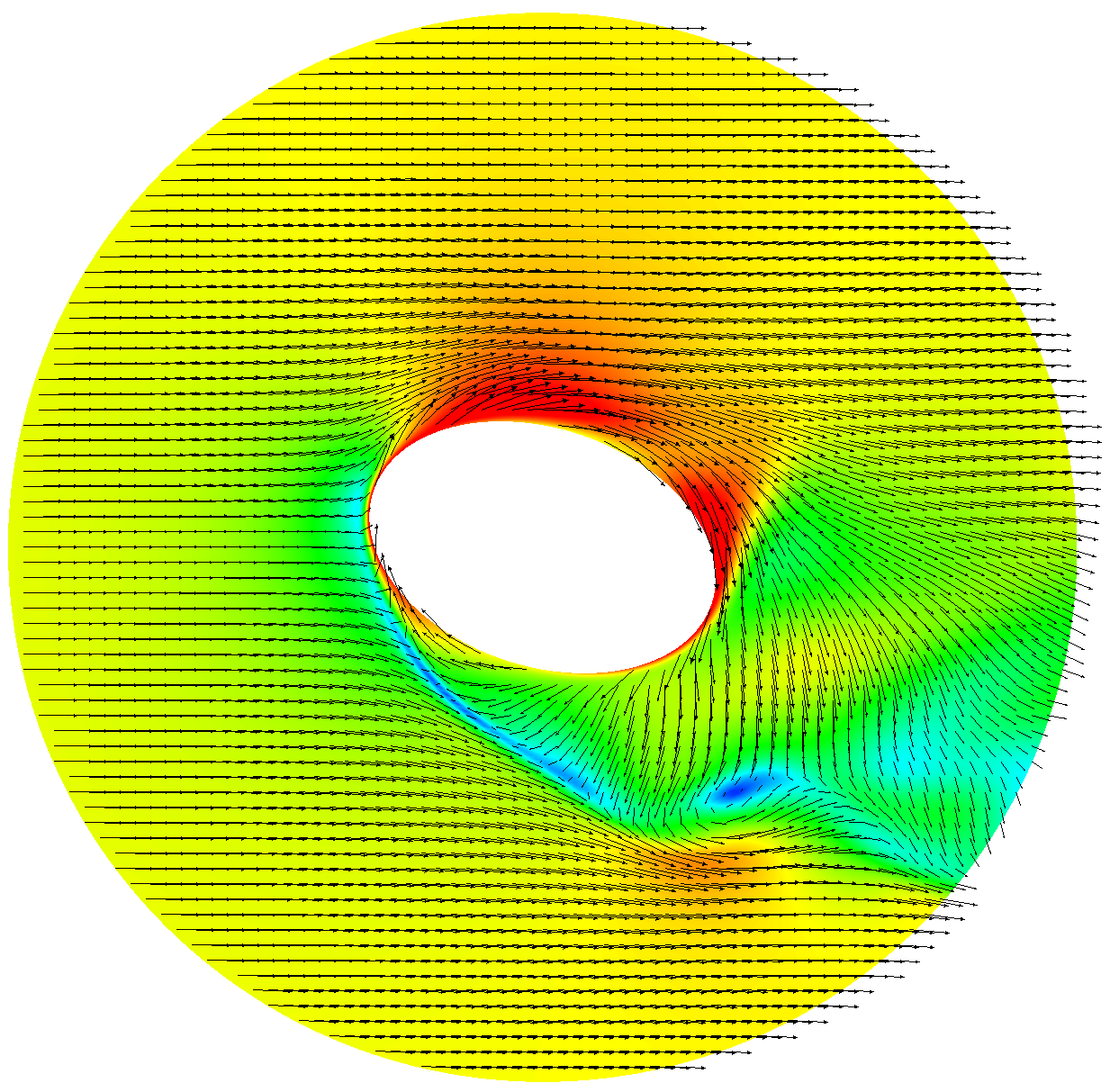} &
                \includegraphics[width=70mm]{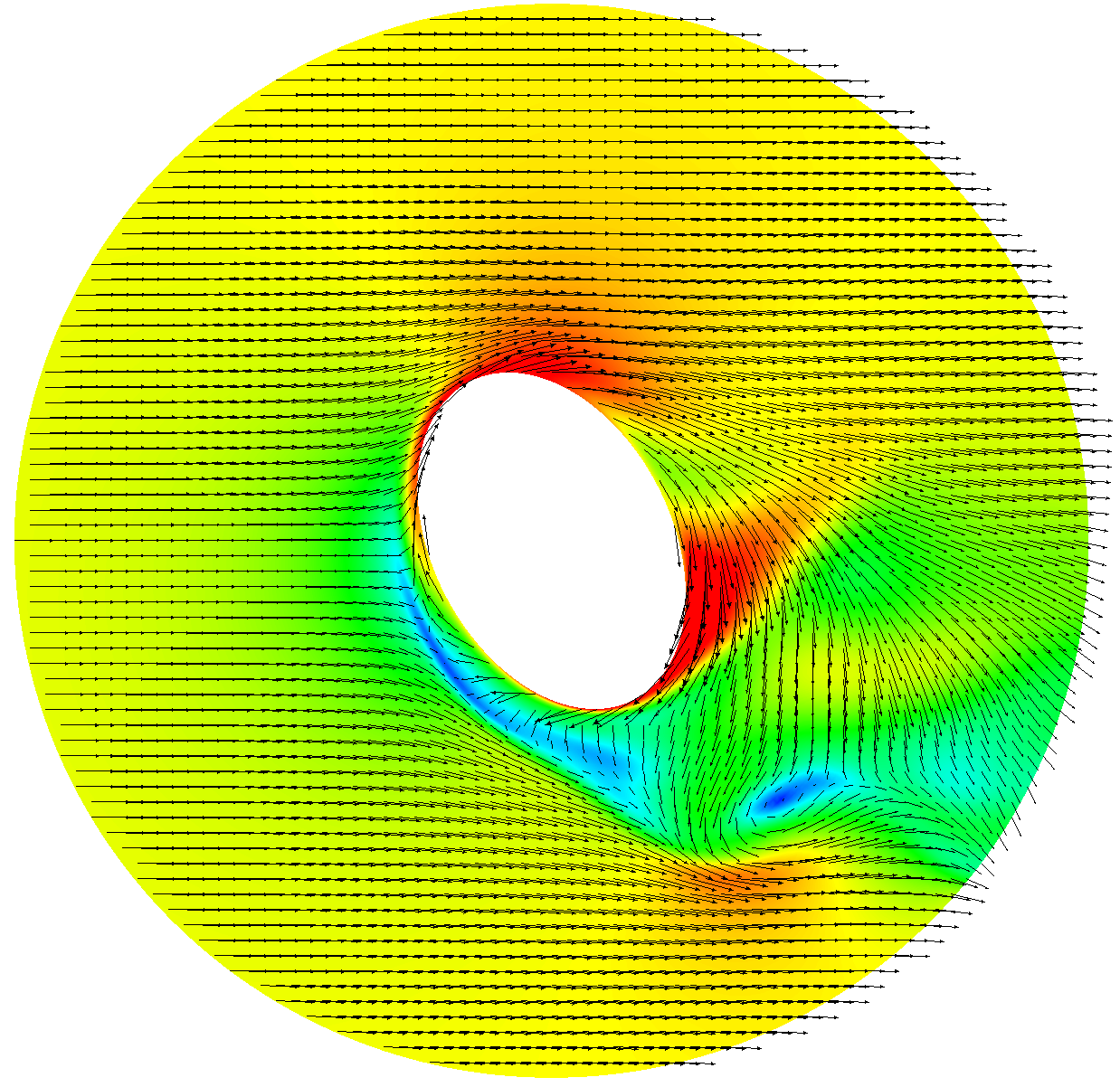} \\
                \textrm{(a) } \theta\approx 108^{\deg} \textrm{(maximum lift)} &
                \textrm{(b) } \theta\approx 153^{\deg} \textrm{(maximum drag)} \\
                \includegraphics[width=70mm]{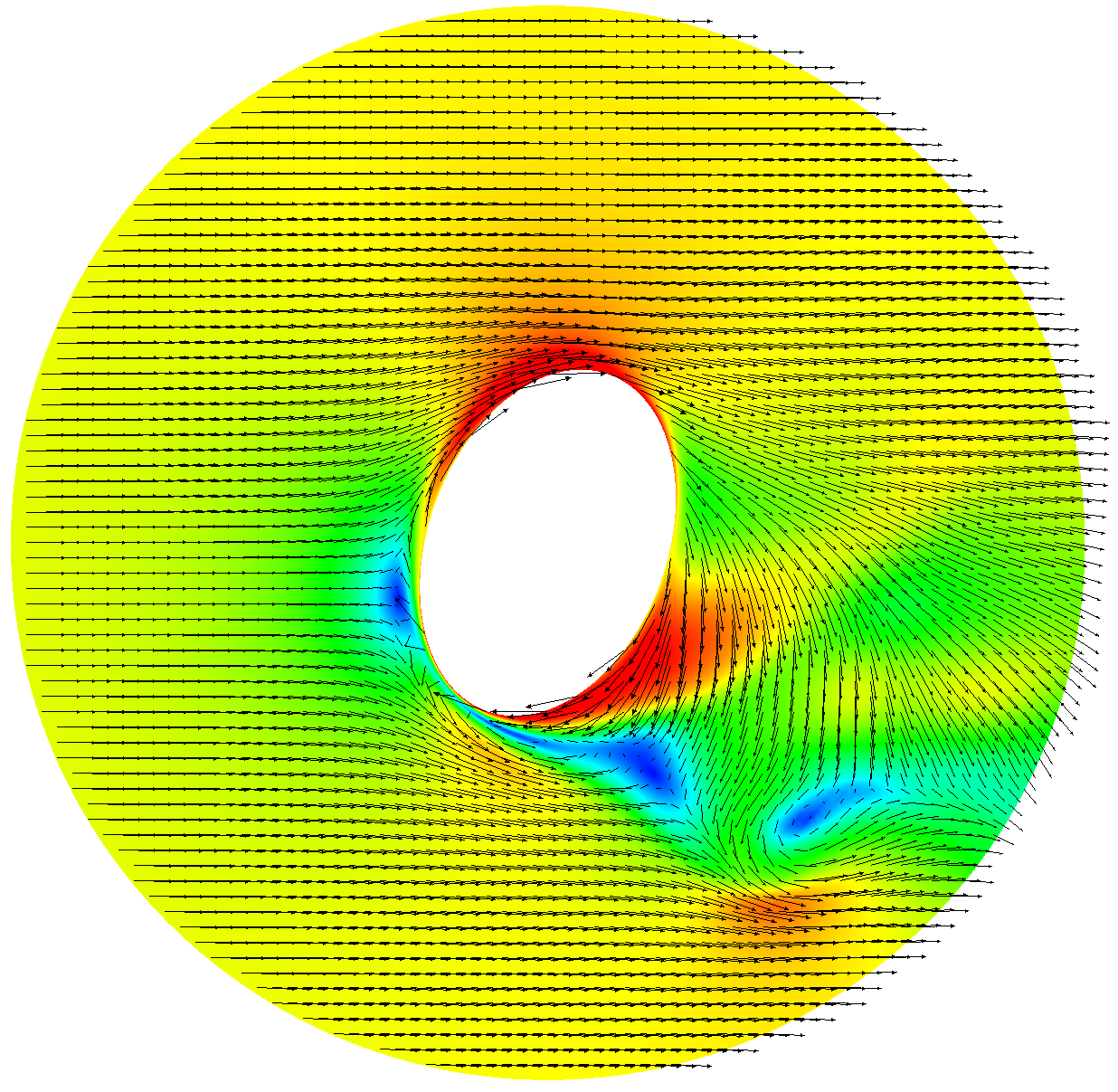} &
                \includegraphics[width=70mm]{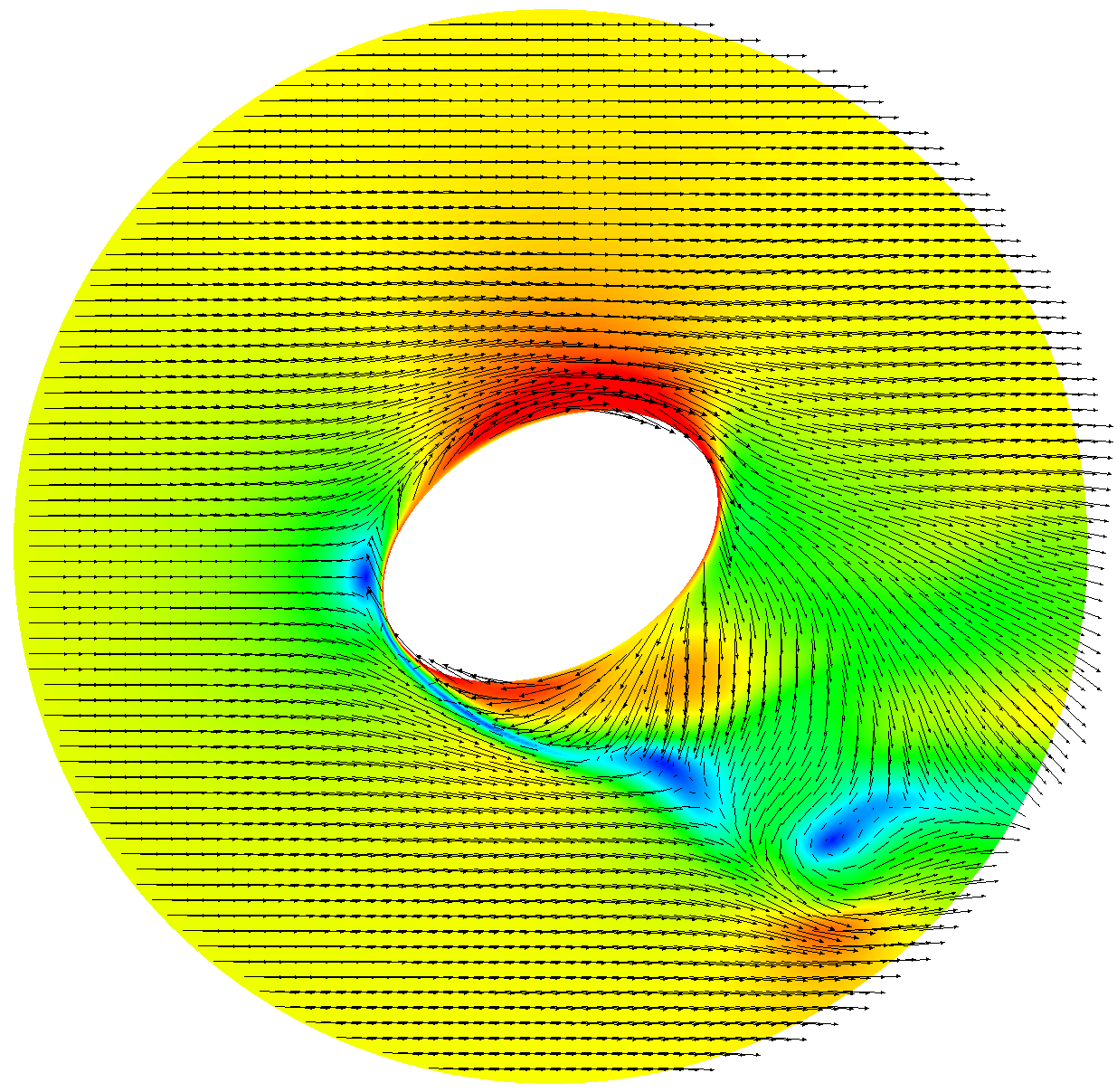} \\
                \textrm{(c) } \theta\approx 200^{\deg} \textrm{(minimum lift)} &
                \textrm{(d) } \theta\approx 242^{\deg} \textrm{(minimum drag)} \\
                \multicolumn{2}{c}{\includegraphics[width=50mm]{velleg}}
                \end{array}$
        \end{center}
        \hspace{-10mm}
        \caption{Slice view of the velocity magnitude with vectors indicating
the velocity field at different orientation ($\theta$) for the nonspherical particle ($a_y=0.75D$).} \label{fig:e1g_dragmax}
\end{figure}

Figures \ref{fig:e1v_dragmax} and \ref{fig:e1p_dragmax} show the variation of
velocity-magnitude and pressure around the ellipsoid for different orientation angles $\theta$, which has been measured clockwise from the axis normal to the flow, as indicated in Fig.  \ref{fig:spheroidschematic}. The results have been plotted on the y-z plane going through its center for $\theta$ ranging from $90^{\deg}$ to $270^{\deg}$.
The nine panels in the figures correspond to orientations of the particle at
maximum ($180^{\deg}$) and minimum ($90^{\deg}, 270^{\deg}$) frontal-area
(projected-area);  and the orientations for maximum lift ($\sim 108^{\deg}$),
maximum drag ($\sim 153^{\deg}$), minimum lift ($\sim 200^{\deg}$), and minimum
drag ($\sim 242^{\deg}$).  Additionally, Fig. \ref{fig:e1g_dragmax} shows the
instantaneous velocity magnitude with vectors indicating the velocity field
for the angles corresponding to maximum and minimum drag and lift forces.

We observe that the rotation of the particle results in the relatively high-speed
flow region, which is attached to the particle surface (indicated by the region in red in Fig.  \ref{fig:e1v_dragmax} and the vector field
in Fig. \ref{fig:e1g_dragmax}),  being pulled from the leeward side to
the windward side of the particle.
The attached flow pulled along the surface of the rotating particle interacts
with the background flow, resulting in the change of high- and low-pressure
zones on the particle surface, when compared with the corresponding non-rotating case.
For example, we observe in Fig. \ref{fig:e1p_dragmax} that as the particle
rotates past $\theta \approx 90^{\deg}$, a low pressure zone is created on the
top surface of the particle due to the background flow going over the rotating
particle. Simultaneously, a high pressure zone forms at the bottom surface of
the particle due to formation of a stagnation point, resulting from the
interaction between the background flow and the opposing flow attached to the
rotating particle's surface.
This flow interaction results in the particle experiencing maximum lift force at $\theta\approx
108^{\deg}$.  Similarly, we observe that the particle experiences maximum drag
force at $\theta\approx 153^{\deg}$ when the pressure difference around the
particle is maximum in the streamwise direction.
In Fig. \ref{fig:e1v_dragmax}-\ref{fig:e1g_dragmax}, we also observe that
due to the shape of the ellipsoid, a steady shear layer never develops behind
the particle, with the flow periodically separating from the bottom.
This flow separation is not present in the case of a rotating sphere, where a
steady layer forms near the surface of the particle.

Based on the results for the ellipsoids with $a_y=0.75D$ and $a_y=0.25D$, the orientation at which maximum drag and lift occurs is a combination
of two factors; variation of dynamic pressure along the surface of the sphere and the
projected area of the particle facing the flow.
For the ellipsoids, this leads to a
phase difference between the orientations at which the drag and lift are
maximum and minimum, and the orientations at which its frontal area is maximum and minimum.
For both the paticles, the phase difference between the maximum-drag and the maximum frontal-area is
$\Phi_D \approx 27^{\deg}$, and the phase difference between maximum-lift and minimum frontal area is $\Phi_L \approx 19^{\deg}$. Additionally, the periodically changing pressure distribution around the particle manifests in such a way that the  orientation at which maximum and minimum drag occurs \emph{leads} the orientation at which maximum and minimum frontal area occurs; whereas the maximum and minimum lift \emph{lags} the orientation at which maximum and minimum projected area occurs. The exact reason why this occurs is a topic that will be explored in detail in the following studies.

In future work, we will continue these analyses for ellipsoidal particles of different aspect ratios and rotation rates to develop a general model that can estimate key flow features
such as maximum, minimum, and mean forces on a particle, and the phase difference between the drag and lift forces and the frontal area.

\subsection{Stationary versus rotating ellipsoid}
\begin{figure}[t] \begin{center}
		$\begin{array}{cc}
		\includegraphics[width=80mm]{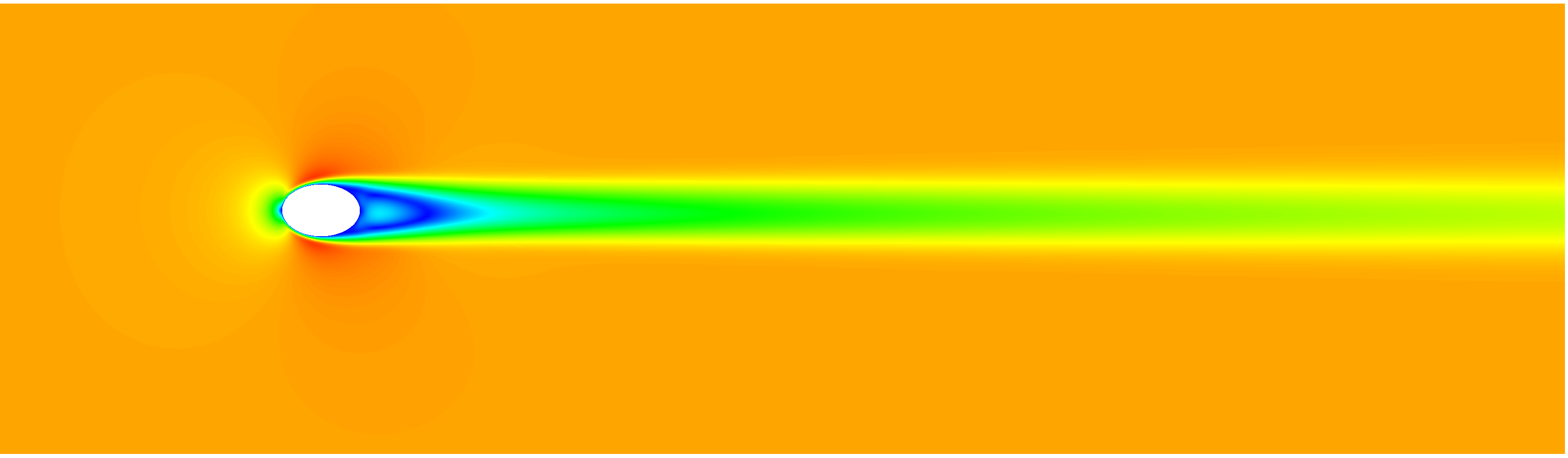} &
		\includegraphics[width=80mm]{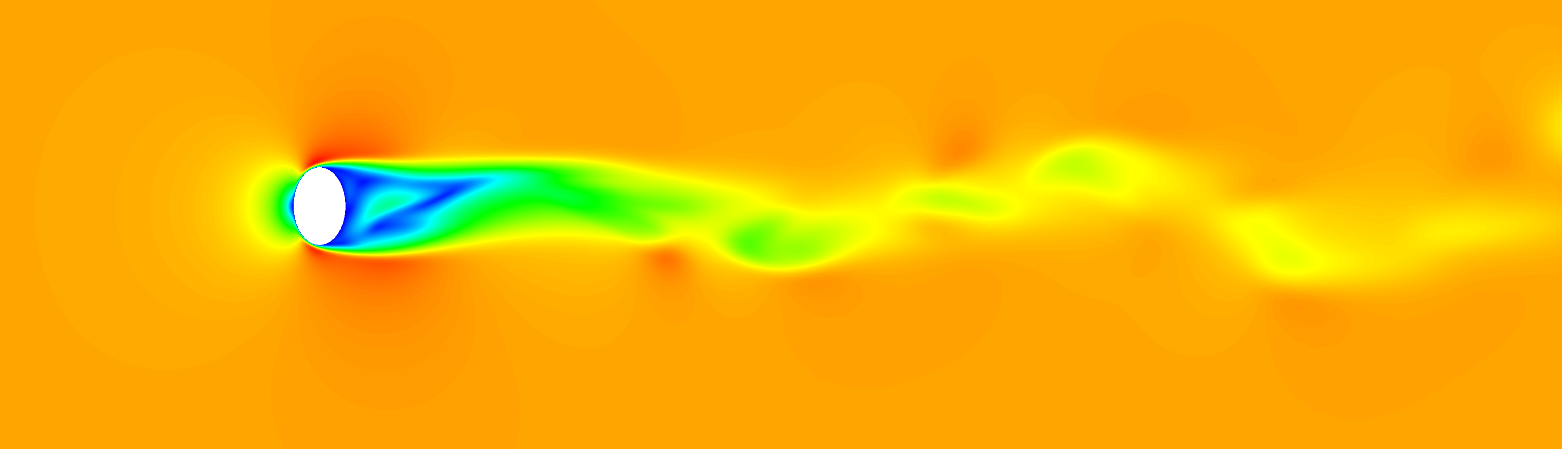} \\
		\textrm{(a) } \theta \approx 90^{\deg} (\textrm{minimum frontal area})  &
		\textrm{(b) } \theta\approx 180^{\deg} (\textrm{maximum frontal area})  \\
		\includegraphics[width=80mm]{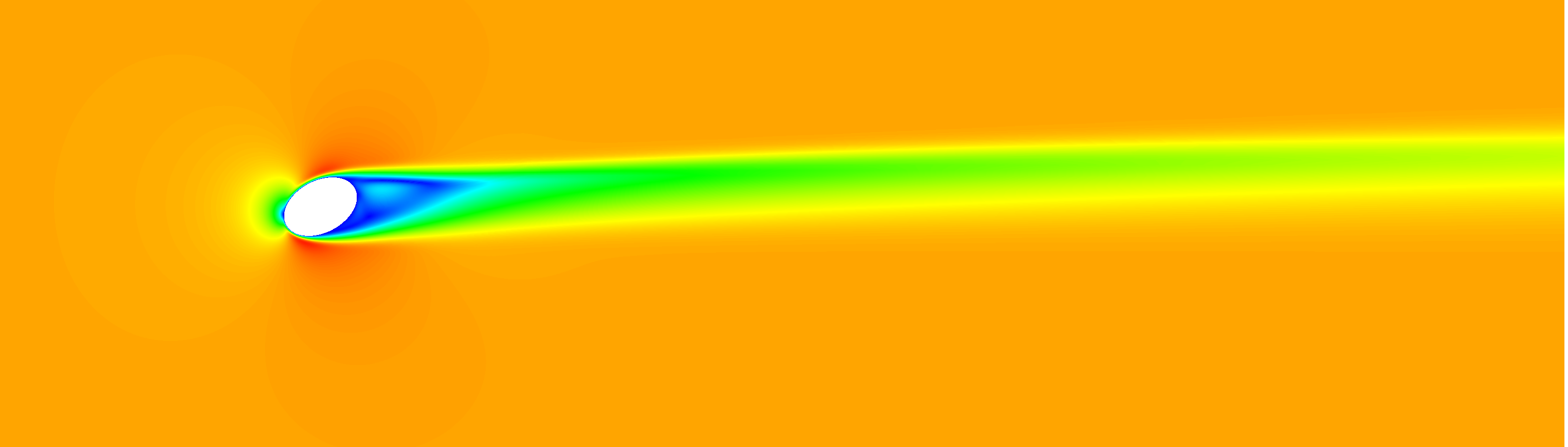} &
		\includegraphics[width=80mm]{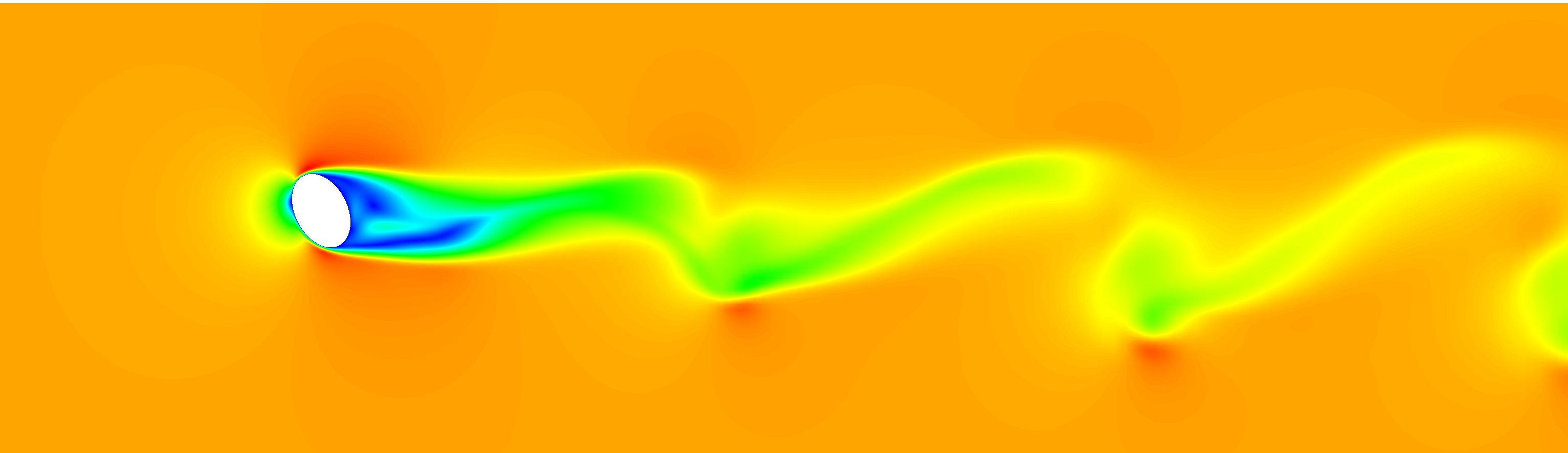} \\
		\textrm{(c) } \theta\approx 242^{\deg} (\textrm{minimum drag}) &
		\textrm{(d) } \theta\approx 153^{\deg} (\textrm{maximum drag}) \\
		\includegraphics[width=80mm]{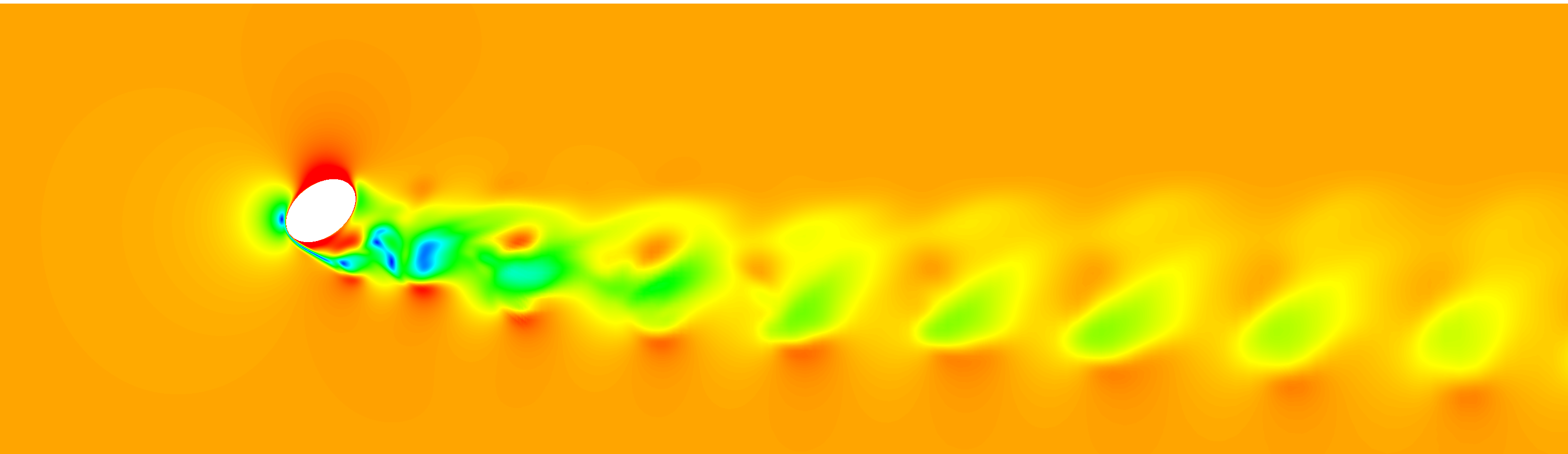} &
		\includegraphics[width=40mm]{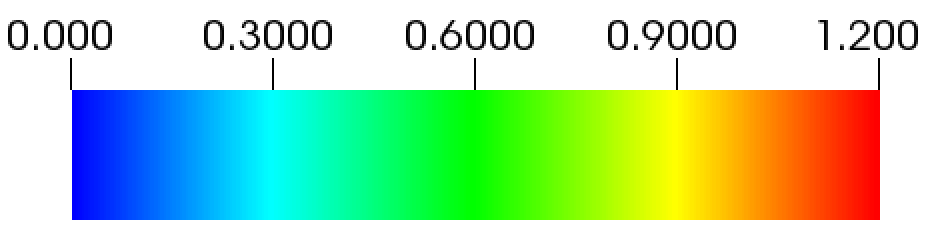} \\
		\textrm{(e) } \theta\approx 235^{\deg} \textrm{ rotating ellipsoid} &  \\
		\end{array}$
	\end{center}
  \vspace{-7mm}
	\caption{Velocity magnitude contours on the y-z plane passing through the center of the ellipsoid
		with $a_y=0.75D$ at different $\theta$. (a) $\theta \approx 90^{\deg}$, (b)
		$\theta \approx 180^{\deg}$, (c) $\theta\approx 242^{\deg}$, and (d)
		$\theta\approx 153^{\deg}$ show the static ellipsoid at different orientation
		angles, and (e) shows the rotating ellipsoid at $\theta\approx 235^{\deg}$ for
		comparison purposes.} \label{fig:e1v_static}
\end{figure}

In this section, we demonstrate that modeling the rotation of the particle is
important for accurately capturing the physics of the flow around it and the
forces experienced by the particle. This is motivated by the fact that in the literature,
there are no relationships for $C_D$ and $C_L$ of ellipsoidal particles that account for the rate of rotation ($\Omega^*$). A survey of the literature shows that existing studies model the effect of rotation either through correlations which have a parameter that accounts for the orientation of the particle \cite{voth2017anisotropic} or using DNS with simulations of static particles at different orientations \cite{ouchene2015,ouchene2016}.
To demonstrate the difference in forces experienced by the particle with and without rotation, we conducted four additional DNS of non-rotating particles at
different orientations, $\theta=90^{\deg}$ (minimum projected area), $153^{\deg}$ (maximum drag), $180^{\deg}$ (maximum projected area), and $242^{\deg}$ (minimum drag), for
the ellipsoid with $a_y=0.75D$, as shown in Fig. \ref{fig:e1v_static}.

\begin{figure}[t!] \begin{center}
		$\begin{array}{ccc}
		\hspace{-10mm}
		\includegraphics[height=50mm]{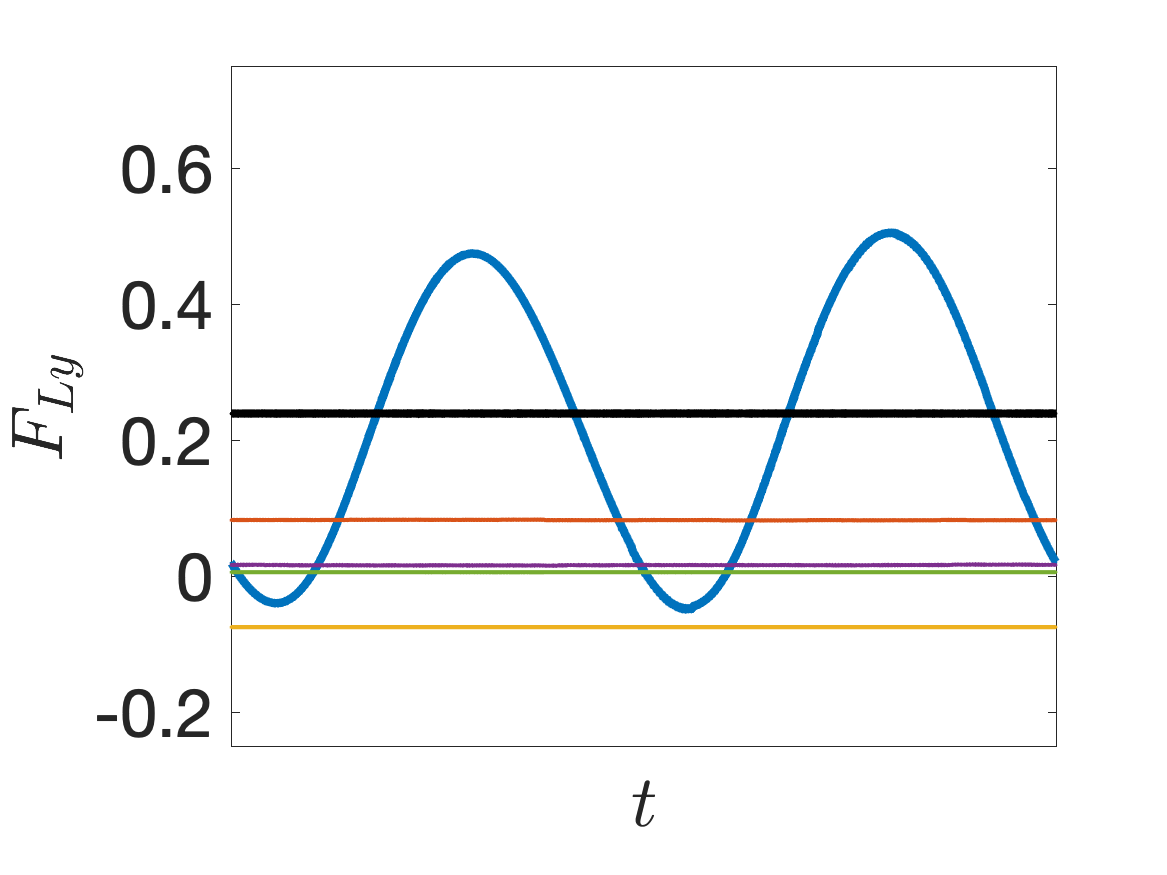} &
		\hspace{-5mm}
		\includegraphics[height=50mm]{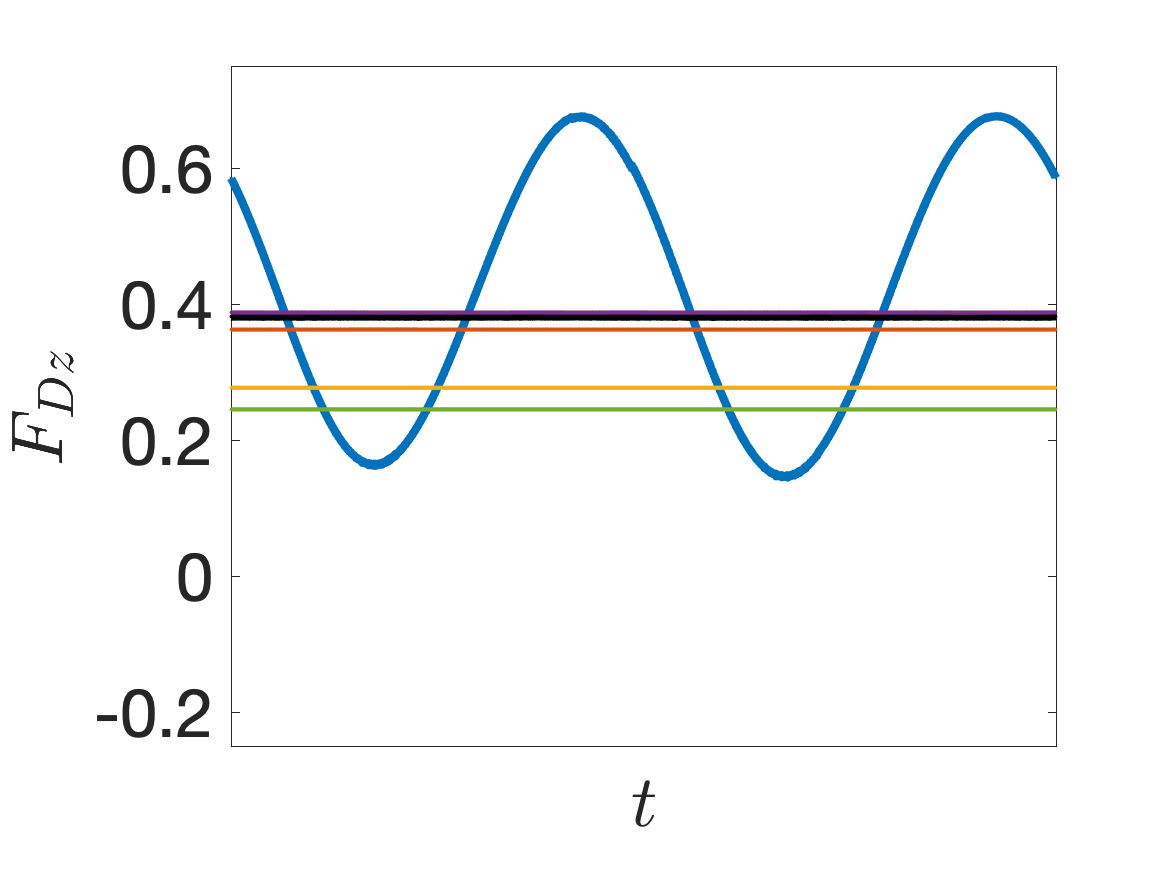} &
		\hspace{-5mm}
		\includegraphics[height=25mm]{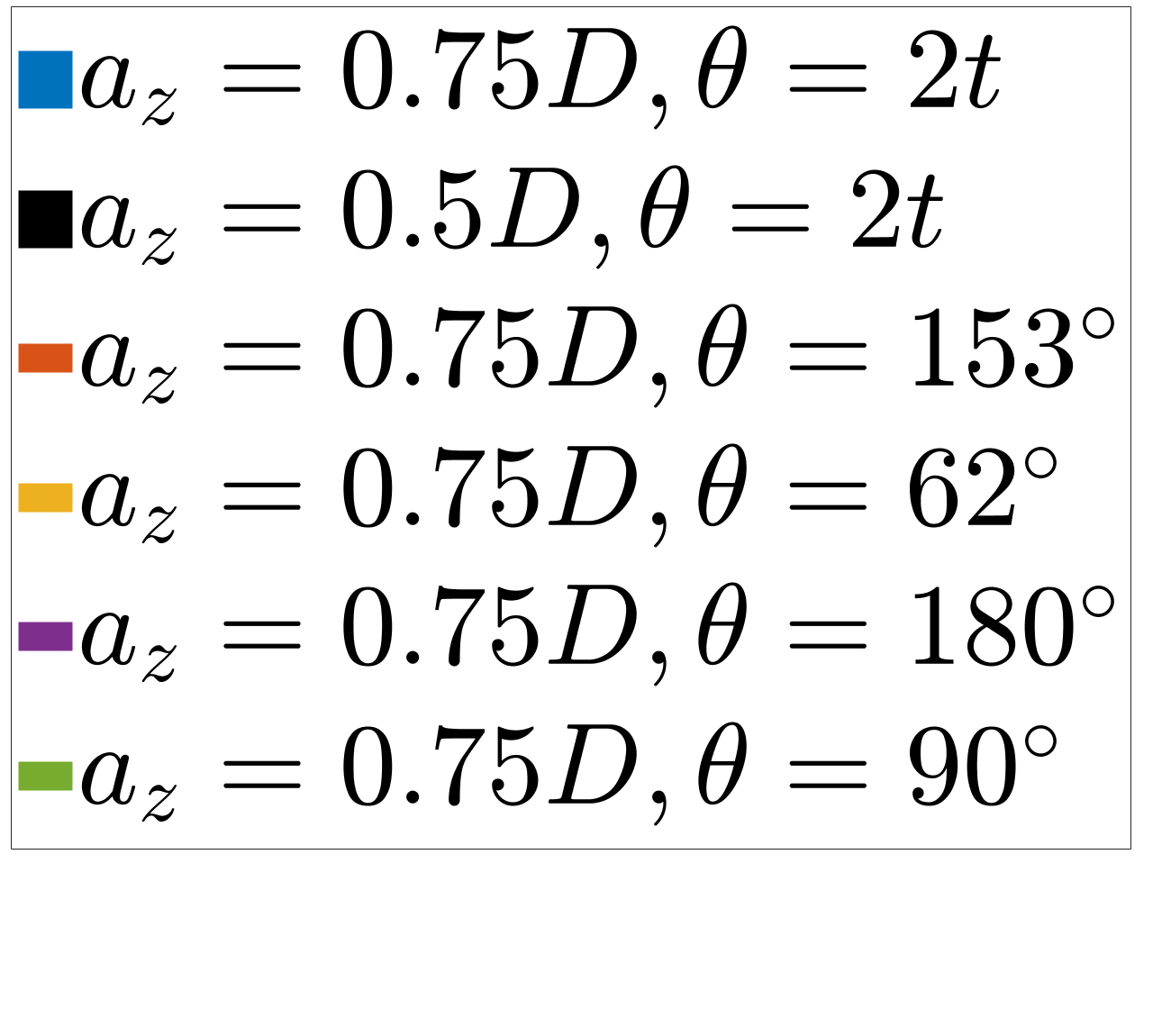} \\
		\end{array}$
	\end{center}
  \vspace{-7mm}
	\caption{Comparison of the (left) lift and (right) drag forces on the static
		ellipsoid at different $\theta$ along with comparison for the forces on rotating
		sphere and ellipsoid.}
	\label{fig:fyfz_static}
\end{figure}

In Fig. \ref{fig:e1v_static}, the velocity magnitude from four non-rotating DNS has been plotted with a snapshot (at $\theta\approx 235^{\deg}$) from the rotating case.
One can observe a clear distinction in the pattern of the flow between the non-rotating cases and the rotating case, even between the two cases with similar particle orientation (Fig. \ref{fig:e1v_static}c and e).
This observation is in agreement with the discussion in the
previous subsection, where we observed the flow repeatedly attaches and detaches from the particle when
it is rotating.
Fig. \ref{fig:e1v_static} shows that neglecting the rotation of the particle does
not accurately model the dynamics of rotating anisotropic particles.

Fig. \ref{fig:fyfz_static} shows the forces acting on the static particles in comparison to the
rotating sphere and ellipsoid ($a_y=0.75D$).
For the static particles, as expected, the particle experiences maximum drag when the projected area
is maximum ($\theta=180^{\deg}$), and the magnitude is not only substantially different from maximum and minimum drag of the rotating case, even the mean drag (ensemble averaged) is about 10\% lower than the
drag experienced for the rotating case.
The effect of rotation is much more obvious on the lift force acting on the particle,
as the lift on the spinning sphere is substantially higher than the lift acting on the non-spinning ellipsoid (almost 2x), even though the ellipsoid has larger projected area than the sphere.

Figures \ref{fig:e1v_static} and \ref{fig:fyfz_static} clearly show that while existing literature could help us
estimate drag or lift coefficients using the forces calculated from the simulations \cite{zastawny2012,dobson2014flow,ouchene2015},
if either the shape or the rotation of the particle is not accounted for,
these correlations cannot be used to accurately predict the motion of anisotropic (ellipsoidal in the current study) particles.

\section{Conclusion}

In this paper, we have presented results from the
ongoing study on flow past rotating particles.
The study is motivated by the need to understand the effect finite sized rotating non-spherical particles have on the flow, and to estimate the resulting forces on the particle. The particles considered in the current study are low-aspect ratio ellipsoids.
Here, DNS was conducted using a highly-scalable implementation of the recently developed  moving nonconforming Schwarz-SEM for three rotating particles (a sphere and two ellipsoids) and for four static ellipsoids at different orientations.
The method used in the current study is capable of scaling to a large number of moving bodies without losing parallel efficiency, and is one of the first high-order methods capable of doing this \cite{mittal2019nonconforming}.

Unlike the rotating sphere,
where the flow is steady near the surface with a shear layer instability
growing downstream of the sphere, the flow for the rotating ellipsoidal particles
experienced recurring separation and reattachment on its surface.
The rotating spherical and ellipsoidal particles were found to bring an attached high-speed flow region from the leeward to the windward side, which results in the \textit{Magnus-Robinson} effect for the sphere \cite{batchelor1967}. For the ellipsoidal particles, this manifests as time-evolving high and low pressure zones around the particle that determines the angle of rotation at which the flow experiences maximum and
minimum drag and lift forces.
Additionally, we observe that changing the shape
of the particle from sphere to an ellipsoid
 leads to a small decrease in the time-averaged lift on the particle.
This happens irrespective of increase or decrease in $a_y$. In contrast, the time-averaged drag was found to increase with increase of $a_y$, and decrease with decrease of $a_y$.
These observations are important because particles are often
modeled as spheres, and the drag and lift coefficients are usually assumed to
be those associated with a sphere.
Finally, the results presented here show that
explicitly modeling a particle's rotation is essential
for accurately capturing the impact of the
particle on the flow and the forces acting on itself.

In future work, we will conduct more DNS simulations to analyze and understand
how shape of the particle, rotation speed, axis of rotation, and the speed of the background flow impact the flow dynamics and forces acting on the particle.
This information will be used to develop correlations for $C_D$ and $C_L$, that accurately account for the effect of shape and rotation of the particle.

\section{Acknowledgments}
This work was supported by the U.S. Department of Energy, Office of Science,
the Office of Advanced Scientific Computing Research, under Contract
DE-AC02-06CH11357.  An award of computer time on Blue Waters was provided by
the National Center for Supercomputing Applications. Blue Waters is a
sustained-petascale HPC and is a joint effort of the University of Illinois at
Urbana-Champaign and its National Center for Supercomputing Applications. The
Blue Waters sustained-petascale computing project is supported by the National
Science Foundation (awards OCI-0725070 and ACI-1238993) and the state of
Illinois. This research also used resources of the Argonne Leadership Computing
Facility, which is a DOE Office of Science User Facility.

\bibliographystyle{elsarticle-num}
\bibliography{lit}

\end{document}